 \documentclass[preprint2]{aastex}

\usepackage{pifont}
\usepackage{latexsym}
\usepackage{txfonts}
\usepackage{ifthen}
\usepackage{url}

\newcommand{\chisq}{$\chi^2$\ }
\newcommand{\see}{\emph{see}\ }
\newcommand{\eg}{\emph{e.g.}\ }
\newcommand{\ie}{\emph{i.e.}\ }

\newcommand{\Qstat}{$Q$--statistics}
\newcommand{\Cstat}{$C$--statistics}
\newcommand{\Ft}[1]{$F$ test at $\alpha = #1 \%$}

\slugcomment{Not to appear in Nonlearned J., 45.}

\shorttitle{Testing tests on AGN microvariability}
\shortauthors{de Diego}

\begin{document}



\title{Testing tests on AGN microvariability}


\author{Jos\'{e} A. de Diego\altaffilmark{1}}
\affil{Instituto de Astronom\'{\i}a, Universidad Nacional Aut\'{o}noma de M\'{e}xico, Ciudad Universitaria, Mexico City 04510, Mexico}





\begin{abstract}
Literature on optical and infrared microvariability in Active Galactic Nuclei (AGNs) reflects a diversity of statistical tests and strategies to detect tiny variations in the lightcurves of these sources. Comparison between the results obtained using different methodologies is difficult, and the pros and cons of each statistical method are often badly understood or even ignored. Even worse, not properly tested methodologies are becoming more and more common, and biased results may be misleading to realize the origin of the AGN microvariability.
This paper intends to point future research on AGN microvariability to the use of powerful and well tested statistical methodologies, providing a reference for choosing the best strategy to obtain unbiased results.
Lightcurves monitoring have been simulated for quasars, reference and comparison stars. Changes for the quasar lightcurves include both Gaussian fluctuations and linear variations.
Simulated lightcurves have been analyzed using \chisq tests, $F$ tests for variances, One--Way Analysis of Variances and \Cstat\ methodologies. Statistical Type\,I and Type\,II errors, which indicate the robustness and the power of the tests, have been obtained in each case.
One--Way Analysis of Variances and \chisq show to be powerful and robust estimators for microvariations, while the \Cstat\ is not a reliable methodology and its use should be avoided.
\end{abstract}





\section{Introduction} \label{sec:introduction}


    Observational techniques for optical monitoring of AGNs and photometric studies of variable stars have many similarities, but they differ in the sense that variable stars often show periodic lightcurve fluctuations, while AGNs do not. Although most variable AGNs (blazars) show microvariability or transient large amplitude variations in a few hours, the unpredictability of these changes in brightness and the difficulty to be confirmed by other observers have been a perennial cause of incredulity and skepticism since the first report by \citet{Mat63} of a 15\,min microvariability event of amplitude $\Delta V = 0.044$ in 3C\,48. In less variable objects, such as quasars, the amplitude of the microvariations are usually lower and close to the limit of detection.

    In order to increase the confidence on the validity of variability reports, a number of statistical tests have been proposed to prove the reliability of the measurements. A methodology that has been widely used is the \chisq test for variances that compares a sample variance obtained from a a suspected variable target with a theoretically calculated variance for a non-variable object, taking into account all the possible sources of error. For example, \citet{Pic83} use a \chisq and the so called \Qstat, based on the difference between the brightest and the dimmer observations, to investigate long term variability in approximately 6000 photographic observations of 130 AGNs of different types monitored during 13 years.
    After the introduction at the end of the seventies of the CCDs in astronomical observations, differential photometry became a very reliable technique for short time resolved lightcurve studies. In differential photometry, the flux of the target object is divided by a \emph{reference star} in the same CCD frame. As the target and the reference star images have been obtained simultaneously at the same air mass and identical instrumental and weather conditions, the flux ratio is considered to be very reliable. It is common practice to compare the differential lightcurves of the target an at least one non variable field star, denoted as \emph{comparison star}.

    A remarkable effort to update statistical techniques used in photoelectric photometer and photographic plate measurements, taking advantage of the better and faster response of CCD detectors, was carried out by \citet{How88}, who compared the variances between the object and a comparison star using the $F$ distribution. The $F$ test compares two sample variances, one for the suspected variable source and the other for the non-variable comparison star. These authors also introduced the $\Gamma$ factor to account for scale differences between the variances of the target and the comparison star due to photon noise. However, \chisq based tests have nevertheless endured as they are well known and simple estimators for flux variations.

    \citet{Jan97} and \citet{Rom99} have proposed a new test to analyze microvariability based on the ratio $C$ between the target and the comparison star standard deviations, rather than variances. This \Cstat\ resembles \citet{Pic83} \Qstat, but is less sensible to the spurious effect of a single discrepant point. In the last decade \Cstat\ has become very popular and several researchers have adopted this statistical methodology to study quasars \citep[\eg][]{Sta04, Gup08} and blazars \citep[\eg][]{Xie04, And05}.

    A different methodology to analyze differential lightcurves, has been proposed by \citet{Die98}. These authors use the One--Way Analysis Of Variance (ANOVA) test for studies of quasar microvariability. Using this technique, \citeauthor{Die98} were the first to claim that microvariability events were as frequent in radio quiet as in radio loud quasars (excluding blazars). ANOVA has also been applied in other studies of AGN variability \citep{Ram04, Vil09, Ram09}. However, because the novel results reported by \citet{Die98} were unexpected at the time and because they were difficult to compare with previous studies, the ANOVA test has still not gained full acceptance \citep{Rom99, Car07}. In this paper, independent runs of data are simulated using Monte~Carlo technique, and analyzed using One--Way ANOVA and other statistical techniques.

    Despite their generic name, which can be misleading, ANOVA tests are designed to detect differences between several sample \emph{means}, rather than between sample \emph{variances}. Thus, such tests can be considered a generalization of the Student \textit{t}-tests for differences between two sample means. It is worth noticing that tests for means can distinguish smaller differences than tests for variances (\see \S\ref{sec:discussion}), and thus it is expected that ANOVA improves the detection of microvariability events. ANOVA tests are used, for instance, in \emph{Experimental Design} statistical methodologies \citep[\eg][]{Box05}, which deliberately impose one or more conditions on different groups of data in the interest of observing the response. These methodologies radically differ from those common in Astronomy, which involve collecting and analyzing data \emph{on the run} and where external conditions cannot be manipulated. Therefore, after applying the same objective method to this and other statistical tests, we will be able to effectively establish the reliability and advantages of the ANOVA-test.

    This paper presents a comparison of the outcome of different analysis strategies for the detection of low amplitude microvariations when reliable astronomical differential photometric data are available. By reliable data it is understood data characterized by random errors that are not affected by systematical errors. Dealing properly with systematical observational errors would require, first, to detect that the data are indeed affected by these errors, second, to derive an understanding of their cause, and third, to properly correct for these systematics by changing and fine tuning the observational set-up, to the extent that this is possible. In this Paper, systematical effects will be entirely disregarded.


    To analyze the data, several procedures based on \chisq tests, $F$ tests for variances, One--Way ANOVA and \Cstat\ will be considered in turns. Even though these consist of common statistical methodologies, there exist many different implementation strategies of these test and, in principle, new alternative tests could be carried out. Strictly speaking, some of the  derived inferences or comparisons established between the various tests may be only valid for the particular cases considered here. Furthermore, we cannot rule out that different observational circumstances or set-up implementations with respect to those envisage in this Paper might result in altogether different results.

    This paper is organized in the following way: the simulation procedure is described in \S\ref{sec:simulations}; results are shown in \S\ref{sec:results}; a discussion and comparisons between the tests are presented in \S\ref{sec:discussion}; and the conclusions are summarized in \S\ref{sec:conclusions}. In \S\ref{sec:test-desc} the interested reader can also find the mathematical description of each test, along some comments on their use and validity. An interesting implementation of the One--Way ANOVA to improve the detection of microvariability in the AGN lightcurves is discussed in \S\ref{sec:improving}.

\setcounter{table}{0}
\begin{table}[t]
 \centering
 \begin{minipage}{0.85\linewidth}
  \caption{Parameters used in the simulations. \label{tab:param}}
  \begin{tabular}{lr@{}lrr}
\tableline\tableline
                & \multicolumn{3}{c}{Electron counts} & Total \\
                \cline{2-4}
   Description  & \multicolumn{2}{c}{Signal}   & Noise & S/N \\
\tableline
 Detector       &  --          &             & 103 & -- \phantom{i} \\
 Sky            &  23,         & 287         & 152 & -- \phantom{i}\\
 $V=17$         &  32,         & 292         & 179 & 126 \\
 $V=15$         & 203,         & 766         & 451 & 418 \\
\tableline
\end{tabular}
\end{minipage}
\end{table}

\section{Lightcurve simulations} \label{sec:simulations}

    In order to accurately analyze the power and robustness\footnote{The power of a statistical test is the probability that the test will reject a false null hypothesis. A robust statistical technique is one that performs well even if its assumptions are somewhat violated by the inherent properties of the sampled population.} of the different statistical methodologies, simulations were performed for the lightcurves of the quasar, a references star and two comparison star. For easiness, the \emph{Instrument Simulator}\footnote{The Instrument Simulator of the OAN has been developed by Alan Watson and can be accessed through the OAN web page \url{http://132.248.4.250/$\sim$resast/simulador}.} software of the Mexican \emph{Observatorio Astron\'{o}mico Nacional} was used to obtain basic data for the simulations. The input arguments were: the 1.5\,m telescope, the SITe1 detector, $2\times2$ binning, $V$ filter, 1.5\arcsec\ seeing, 3\arcsec\ aperture, 60\,s exposure time, and magnitudes $V=17$ and 15 for the quasar and the reference star, respectively. For simplicity, the comparison stars have been chosen to have the same magnitude as the quasar. The output includes, among others, the following parameters: the signal to noise ratio, and the object, sky and detector (total read-out) noises. From these parameters, the object and the sky electron counts were obtained considering Poisson distribution (\ie photon shot noise). The parameters used in the simulations are shown in Table \ref{tab:param}. Column (1) describes the source associated to the parameters; columns (2) and (3) indicate the electron counts for the signal and noise, respectively; column (4) shows the signal noise ratio (SNR).

\begin{figure}[t]
%
 \ifthenelse{\boolean{@twocolumn}}{\epsscale{2}}{\epsscale{1}}
 \plottwo{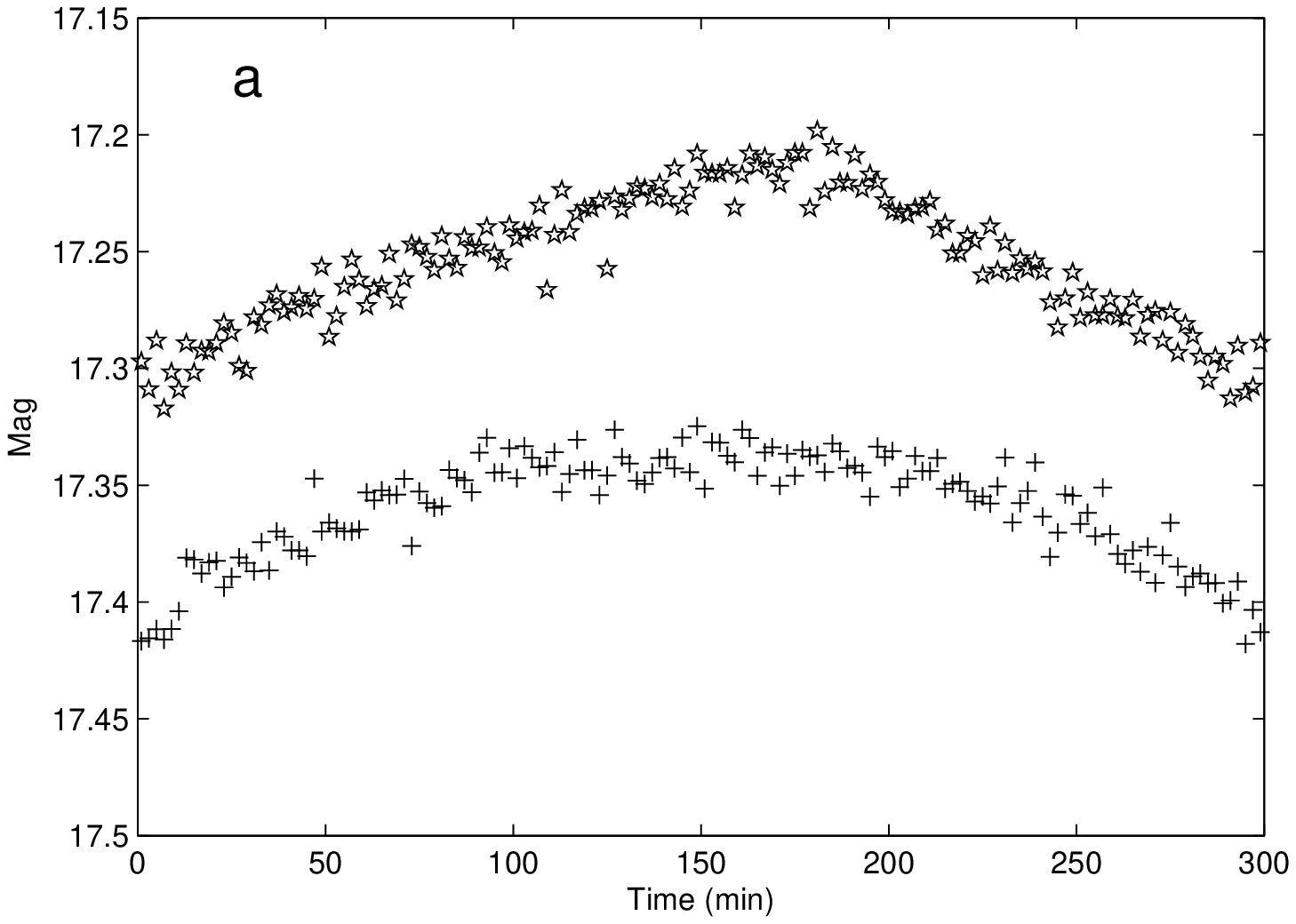}{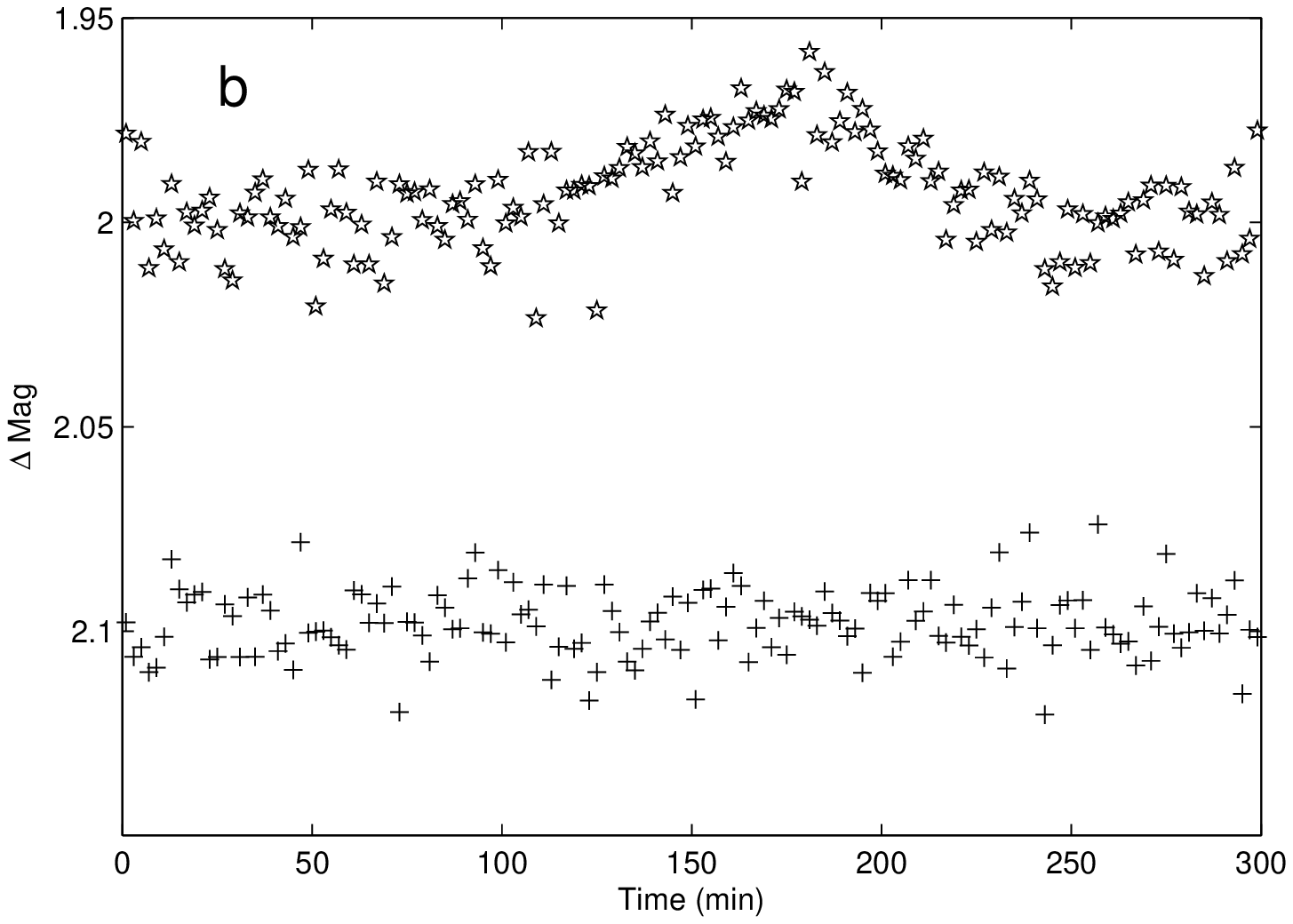}
 \caption{Simulated raw curves (a) and Differential lightcurves (b) for $V=17$ quasar (\ding{73}) and comparison star ($+$, with an offset of 0.1\,mag). At the middle of the simulation, the objects are crossing the zenith, where the atmospheric absorption has been set at 20\%. A reference star with $V=15$ has been used to calibrate the observations. The amplitude of the variability in this case was 0.0267, and the peak has a FWHM of 60\,min.} \label{fig:lc_gauss}
\end{figure}

\begin{figure}[t]
 \ifthenelse{\boolean{@twocolumn}}{\epsscale{2}}{\epsscale{1}}
 \plottwo{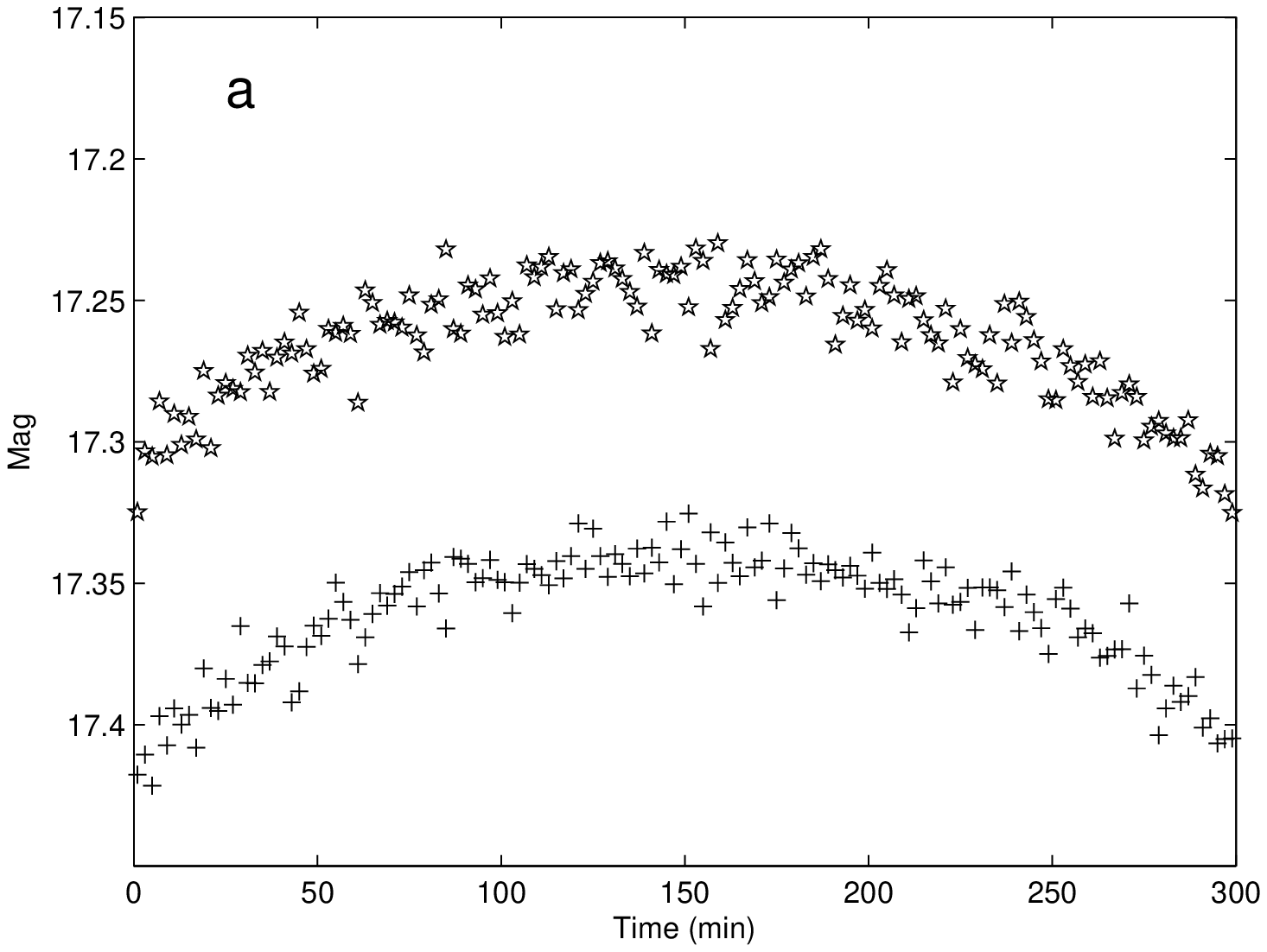}{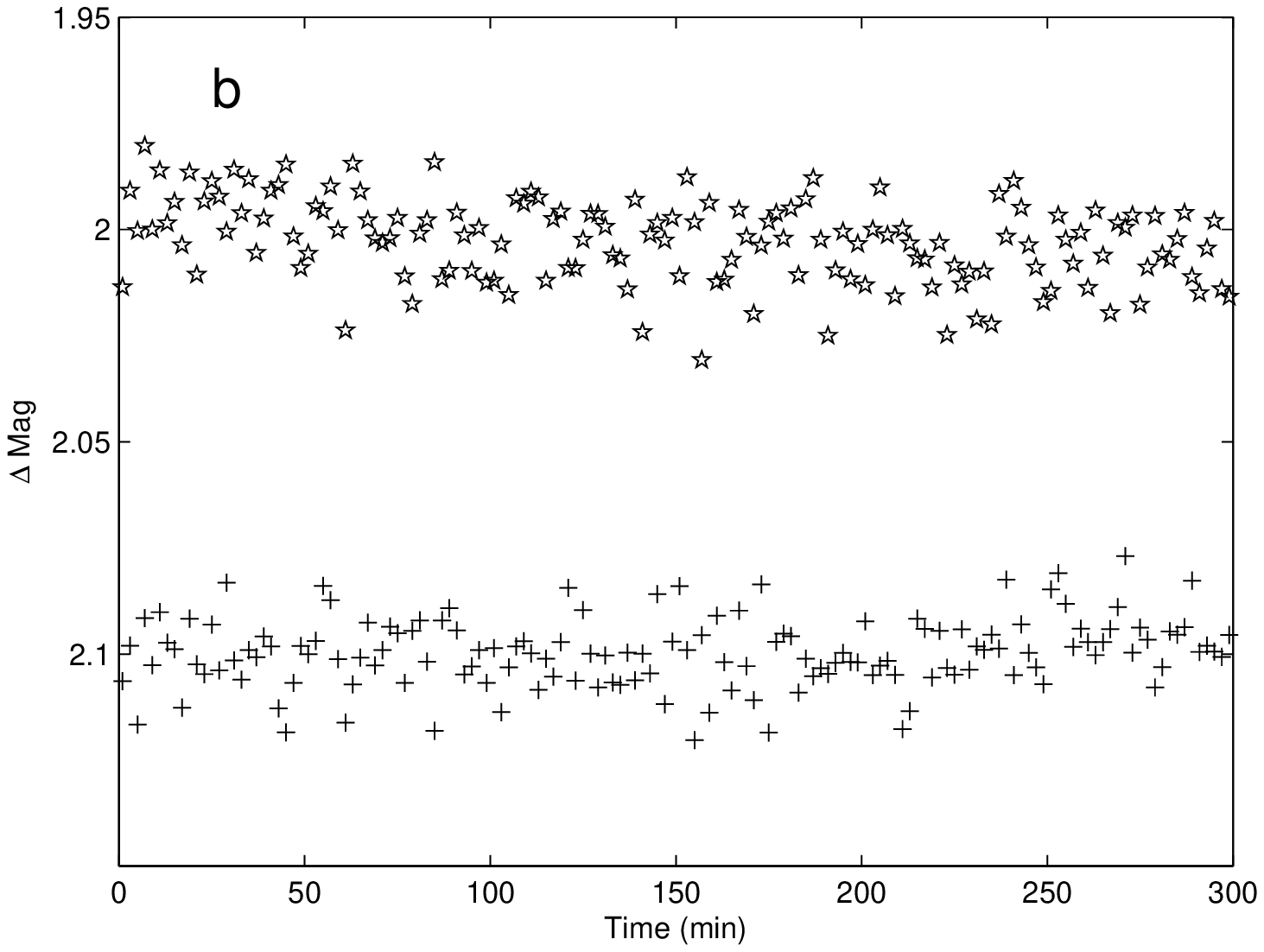}
 \caption{Simulated raw curves (a) and Differential lightcurves (b) for $V=17$ quasar (\ding{73}) and comparison star ($+$, with an offset of 0.1\,mag). At the middle of the simulation, the objects are crossing the zenith, where the atmospheric absorption has been set at 20\%. A reference star with $V=15$ has been used to calibrate the observations. The amplitude of the variability in this case was 0.0113, and it has a steady increase in flux during the monitoring.} \label{fig:lc_linear}
\end{figure}

    For a telescope of 1.5 and a fairly SNR larger than 100, 1\,min exposures are reasonable. Longer exposures may saturate bright objects and stars which might be used as reference in differential photometry. Thus, every simulation comprises $150 \times 1\,\textrm{min}$ exposures during 5\,h of monitoring, with 1\,min lag between exposures to account for CCD read--out. For more realistic simulations, atmospheric attenuation in the $V$ band ($A_V = 0.2 \, \textrm{mag/air mass}$) has been taken into account. The object is supposed to pass through the zenith in the middle of the run (\ie 2.5\,h after the monitoring began). Photon noises were generated for the object, the reference and comparison stars, as well as for a constant brightness sky. Finally, white gaussian noise was also generated to account for the CCD read-out. The photometric accuracy for the differential lightcurves of the quasar and the comparison star is about 0.01\,mag.

\begin{table*}
 \centering
 \begin{minipage}{.9\textwidth}
  \caption{Results of the simulations.} \label{tab:results}
  \begin{tabular}{lrrrrlrrr}
  \tableline\tableline
                    &    & \multicolumn{3}{c}{Gaussian peak variation} & & \multicolumn{3}{c}{Linear variation} \\
                    \cline{3-5} \cline{7-9}
 Test               & \multicolumn{1}{c}{$\alpha$}  & Type\,I   & Detections    & Type\,II  & & Type\,I    & Detections    & Type\,II \\
 \tableline
 \chisq test        & 0.001         &    10     & 1614          & 4386      & &     2          & 2278          & 3722 \\
 {   }Idem (std)    & 0.001         &   125     & 1748          & 4252      & &   129        & 2385          & 3615 \\
 One--Way ANOVA     & 0.001         &     3     & 2187          & 3813      & &     4          & 3043          & 2957 \\
 {   }Idem (30 min lags) & 0.001         &     2     & 1086          & 4914      & &    11         & 1668          & 4332 \\
 $F$ test ($\alpha=0.1\%$)   & 0.001         &     9     &  837      & 5163      & &     5          & 1302          & 4698 \\
 {   }Idem ($\alpha=1\%$)    & 0.01          &    63     & 1443      & 4557      & &     61         & 2050          & 3950 \\
 \Cstat             & 0.01          &     0     &    0          & 6000      & &     0          &    0          & 6000 \\
 \tableline
\end{tabular}
\end{minipage}
\end{table*}

    Two runs of simulations were performed. One of the runs considered a gaussian shaped variation in the quasar flux. The duration of the variations are described by their FWHM that are allowed 60 discrete values in the range between 1 and 60~min (\ie in steps of 1~min). This range for the duration of the variations is also appropriate to investigate the possibility of detecting spikes \citep[\see][]{Sag96, Die98, Gop00, Sta04}. The peak of the variation was allowed to be centered between 120 and 180~min from the beginning of the monitoring. This range for the peak center ensures that the whole variation is contained in the data set. Finally, for each of the 60 values set for the duration of the variation, 100 simulations were performed allowing random amplitudes up to 3\% of the quasar flux (\ie $\sim 0.03\,\textrm{mag}$). Therefore, the total number of simulations were 6000 ($60 \times 100$), of 150 photometric points each. Fig.~\ref{fig:lc_gauss}a shows the final simulated raw curve for the quasar and comparison star for a Gaussian variation, while Fig.~\ref{fig:lc_gauss}b shows the same curves after calibration by the observations of the reference star. The standard deviation of the differential curve of the comparison stars shows that the photometry is accurate up to 0.009\,mag.

    For the other run, the lightcurves of the objects present a constant (linear) flux variation, as shown in Fig.~\ref{fig:lc_linear}. The amplitudes of these variations, measured as the difference between the first and the last data point, were also random valued up to 3\% of the quasar flux. As in the gaussian peak case, a total number of 6000 simulations of 150 data points each were performed.
    In both cases, Gaussian peak and linear variation, all the fluxes and estimated errors were converted into magnitudes for the analysis.

\section{Results} \label{sec:results}

    The results of the statistical analysis of the simulations are summarized in Table~\ref{tab:results}. Column (1) identifies the test (\see below); column (2) indicates the significance level $\alpha$; columns (3), (4) and (5) show the number of Type\,I errors found in the analysis of one of the comparison stars, the number of detections of microvariations in the quasar lightcurves, and the number of Type\,II errors for the quasars, respectively, for the 6000 simulations for the Gaussian peak variation; columns (6), (7) and (8) repeat the same numbers but for linear variations.

    The first item in Table~\ref{tab:results} is the \chisq test described in \S\ref{sec:chi2-test}. To perform this test, the true error distributions introduced in the simulations (photon and white gaussian noises) have been considered to estimate the error of each individual data point. In the second test, a similar \chisq analysis has been performed, but considering the standard deviation of the comparison star instead of the individual errors of each measurement (note that in this case Type\,I errors have been calculated from the other comparison star). The third test is One--Way ANOVA, performed by grouping the data in sets of 5 individual observations (\see description in \S\ref{sec:anova-test}). The fourth test is also One--Way ANOVA but with 30~min lag between group sampling (and thus it is the only test that considers only a fraction of 1/3 of the simulations). The results of two $F$ tests for variances (\see \S\ref{sec:F-test} for a description) are reported in the fifth and the sixth items, the former at the significance level $\alpha = 0.001$  (or 0.1\%) to compare with the previous tests, and the later at $\alpha =0.01$ (or 1\%) to compare with the \Cstat. Finally, the fifth test is \Cstat\ as described in \S\ref{sec:C-test};

    The significance level $\alpha$ is a probability set \emph{a priori} by the researcher that a test yields, only by chance, a result at least as extreme as the one observed. Note that ANOVA and \chisq tests have been performed for a significance level of 0.1\%, that corresponds to the usual detection limit of $3\sigma$. On the other hand, the significance level of the \Cstat\ is set at 1\%, or a detection limit of $2.576\sigma$, which is the level commonly defined for this test \citep[\eg][]{Jan97, Rom99}. For the $F$ test, both significance level values are considered: $\alpha=0.1\%$ to compare with the \chisq and ANOVA tests, and $\alpha=1\%$ to compare with the \Cstat, as the $F$ test is considered as an alternative to this methodology (\see \S\ref{sec:C-test}).

    Type\,I errors are due to the rejection of a true null hypothesis (\ie rejecting the non-variability hypothesis for a non-variable object). For an unbiased test, the actual number of Type\,I errors depends only on the number of data sets examined and the significance level of the test. In this case, the number of Type\,I errors has been obtained testing the simulated differential curves of a (non-variable) comparison star. For a significance level of say 0.1\% and 6000 simulations, we should expect around 6 spurious detections. Considering a Binomial distribution for the number of Type\,I errors, we expect that its actual number for a given test will be between 0 and 13, and in most cases in the $6 \pm 3$ interval. If the number of Type\,I errors for the comparison star is significantly different from the expected frequencies, it is evidence that the actual significance of the test differs from its nominal set value and the test is not reliable.

    The number of detections in Table~\ref{tab:results} indicates how many tests have succeeded in detecting variability in the quasar simulated differential lightcurves, and it is also a measure of the power of the test, which generally varies as a function of the data set characteristics. On the other hand, Type\,II errors are due to the acceptance of a false null hypothesis (\ie accepting the non-variability hypothesis for a variable object). Note that in \emph{all} the simulations the quasar varied. Therefore, as the number of Type\,I errors is low with respect to the number of detections, the number of Type\,II errors would be (approximately) 6000 less the number of detections.\footnote{A more accurate calculation would take into account the fraction of Type\,I errors included in the number of detections. If Type\,I errors were frequent, mixtures of honest detections and Type\,I errors should also be present.}


    From Table~\ref{tab:results} we see that the \chisq test performed taken into account the actual error distribution, One-Way ANOVA, One-Way ANOVA with 30\,min lags, and the \Ft{0.1}, all show a number of Type\,I errors in accordance with expectations. The \Ft{1} and the \Cstat\ have a lower significance level and accordingly the number of expected Type\,I errors would be 60. The results for the \Ft{1} agree with this expectation, nevertheless \Cstat\ consistently produced neither Type\,I errors nor detections, in accordance with the arguments presented in \S\ref{sec:C-test}.
    In fact, in these simulations, \Cstat\ is always a factor $\gtrsim2$ below its critical value (2.576) that determines the boundary between rejecting or accepting the null hypothesis.

    The \chisq test, performed considering the standard variation of the comparison star instead of the actual errors for each observation in the quasar lightcurve, shows a number of Type\,I errors much larger than expected (Table~\ref{tab:results}). This is consequence of employing a \emph{wrong} methodology. As explained in \S\ref{sec:C-test}, not considering the number of degrees of freedom in the estimation of the standard variation of the lightcurve of the comparison star produces a biased statistics. The increase in the number of detections with respect to the previous \chisq test is not significant. As the number of Type\,I errors is still much less than the number of Type\,II errors, it is possible to estimate the number of spurious detections of variability in the quasar lightcurves (\ie detections not related to the actual variation of the lightcurves). In this case the number of spurious detections will be approximately $125 \times 4252 / 6000 = 89$, while a similar calculation produces 7 spurious detection for the previous \chisq test. Thus, even if the biased statistics would affect only the number of Type\,I errors and not Type\,II errors (which probably will be also affected), we would expect a difference of about 80 detections between both tests, counting for most of its factual value of 134.

\begin{figure}[t]
 \epsscale{0.95}
 \plotone{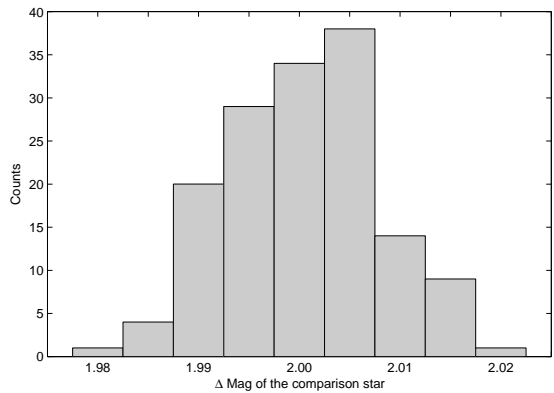}
 \caption{Histogram of the differential lightcurve of a comparison star. In some cases, as in this simulation which shows a strong skewness, the distribution of the differential lightcurves separate from a Gaussian curve.} \label{fig:histog}
\end{figure}

    On the contrary, for the $F$ tests the number of Type\,I errors agrees with the expected number, as commented above. However they show a rather small number of detections in comparison with the other tests (\see Table~\ref{tab:results}). Thus, the detections for the \Ft{0.1}\ are well below the results for the One--Way ANOVA with 30\,min lags, even if the number of data points for the later test is a factor 3 smaller than for the $F$ test. In the case of the \Ft{1}, the significance level is low enough that a larger number of detections would be expected with respect, for example, to the \chisq test at a significance level of 0.1\%; yet the opposite is true. There are two possible explanations for these results, one is the non-robustness of the $F$ test for non-Gaussian distributed data \citep[\S5.4]{Leh86}, and the other is that the test has intrinsically less power than \chisq and ANOVA. Indeed, some of the comparison star differential lightcurves are not well fitted using a Gaussian profile, as shown in Fig.~\ref{fig:histog}, perhaps as a consequence of the underlying Poisson distribution associated with flux measurements. To investigate this possibility, a set of Kolmogorov-Smirnov for Goodness-of-Fit tests of Gaussianity \citep[\eg][\S5.3.2]{Wal03} was performed on the distribution of the simulated Gaussian Peak data for each of the 6000 lightcurves of the comparison star. The significance level of the test was fixed at 5\%; therefore it was expected around 300 Type\,I errors if the data was fairly Gaussian distributed. The actual number of differences found by the Kolmogorov-Smirnov test were 296, which rules out the non-Gaussian distribution explanation. Then, the differences in detecting lightcurve variations with the results of the \chisq and ANOVA test should be imputed to a relative diminished power of the $F$ test.

    The number of detections of the One-Way ANOVA tests is significantly larger than the \chisq tests. The 150 observations computed for each run of simulations are divided in 30 groups of 5 observations each, and the time interval in each group expands 10\,min (remember that exposures and read-outs last 1\,min each). Even if the number of groups is reduced to 10 (\ie sampling an object every 30\,min as in the case of One--Way ANOVA with 30\,min lags), ANOVA maintains its robustness and to a large extent the power to detect microvariations.

\section{Discussion} \label{sec:discussion}


    To perform the ANOVA test it is necessary to bin the data. Although the results for the \chisq test using the standard deviation of the comparison star shows the risks of introducing what might be considered \emph{a priori} reasonable changes in the test strict procedure, it is still questionable if the \chisq test will perform as well as ANOVA if data were binned in the same way. However, just as there is a loss of information in going from a list of observations to an histogram, the binning procedure applied to the \chisq test will produce an immediate loss in the test power. Statistical theoretical backgrounds to reject the \chisq binning methodology may be traced back at least to lossy compression methods considered in Shannon's information theory. Binning data has the effect of reducing the signal besides the noise, and it is mathematically impossible to get any additional information from the binned data. Thus, analyzing both the raw signal and the binned data with the same \chisq statistical procedure will result in a unavoidable loss of sensitivity of the test.

    Things get even worse when the signal is low, close to the 3 sigma limit as it is frequent in microvariability studies. Besides blurring the signal, binning data in groups of $n$ observations in a \chisq procedure has also the effect of reducing by a dividing factor of $n$ the degrees of freedom of the statistical analysis. Both combined effects are disastrous in order to detect tiny variations. For the simulations presented in this paper, the raw data \chisq procedure was able to detect 1614 variations of a total of 6000 cases, while an aside calculation based on the binned procedure for $n=5$ yielded only 284 detections. In comparison, ANOVA produced 2187 detections. Although ANOVA also groups the data producing a loss of signal, it redistributes de degrees of freedom between groups and error estimates within groups, rather than canceling them. Thus, if $N$ is the number of observations and $k$ the number of groups ($k = N/n$), the degrees of freedom are $\nu_1 = k-1$ for the groups and $\nu_2 = N-k$ for the errors, and therefore $\nu_1 + \nu_2 = N-1$, that corresponds to the degrees of freedom of the original dataset. Besides, as stated above,  ANOVA tests (group) means, while \chisq tests variances, and it is well known that tests for means are more powerful than tests for variances, among other things, because the actual value for means are tighter constrained. This is a result of the squaring of each term, which effectively weights outliers and large errors more heavily than small ones. For example, a few simulations generating samples of size 100 drawn from a normally distributed population with mean $\mu=0$ and standard deviation $\sigma=1$ show that the ratio between their respective 95\% confidence intervals is $\mathrm{C.I.}(\sigma^2)/\mathrm{C.I.}(\mu) \approx 1.5$.

\begin{figure}[t]
 \epsscale{0.90}
 \plotone{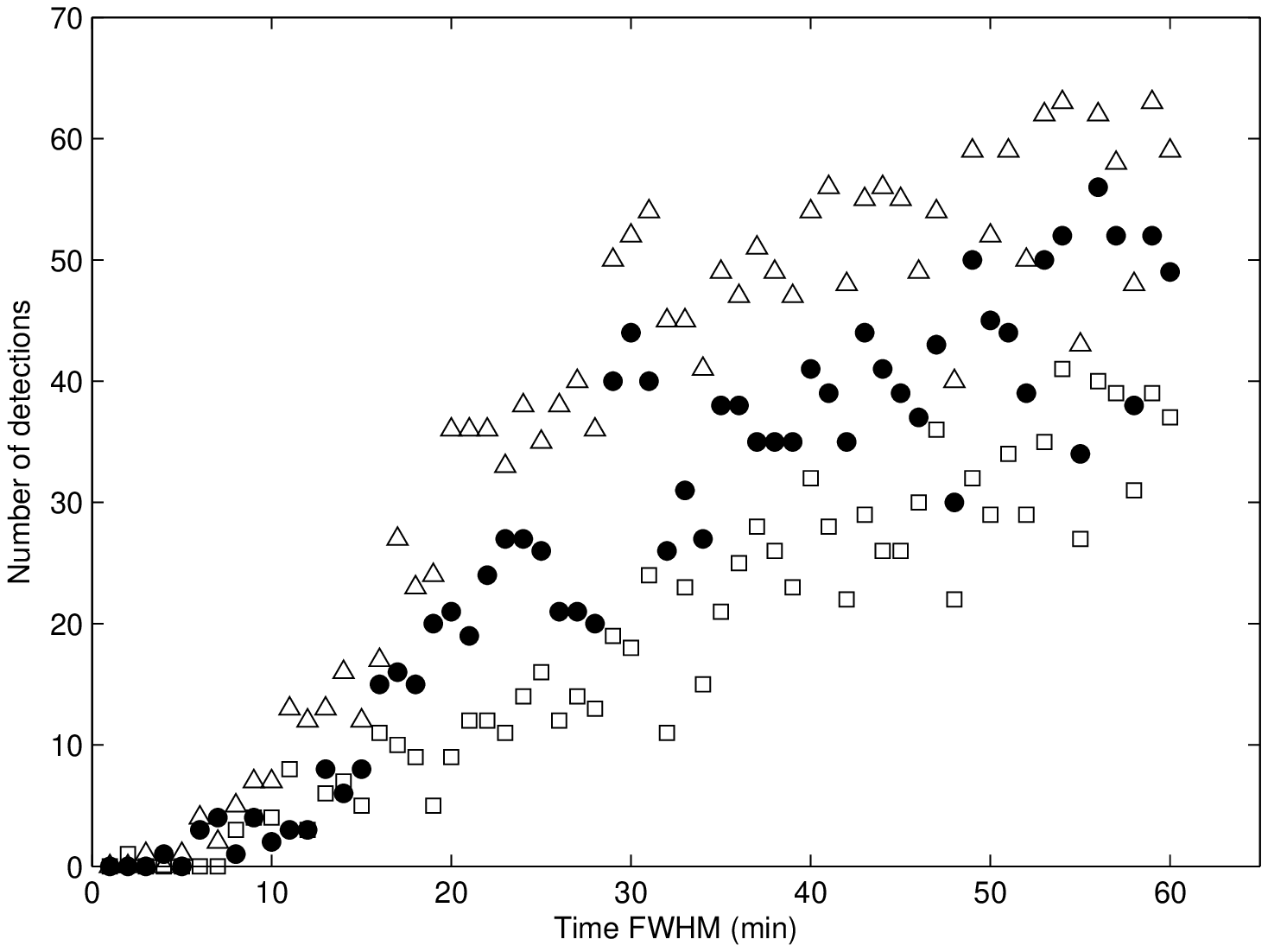}
  \caption{Detection frequencies against temporal FWHM of the variation for the Gaussian peaks. \chisq test ($\medbullet$), One--Way ANOVA ($\triangle$), and One--Way ANOVA with 30\,min lags ($\Box$).}\label{fig:det-vs-dur}
\end{figure}

    In the case of ANOVA and ANOVA with 30 min lags, the time interval within each group has been chosen such that it does not exceed 10 min. This limit is imposed by previous experience that optical microvariations in timescales of less than 20\,min in quasars are rare, hard to detect, or both. On the other hand, when the monitoring is performed in several optical bands that will be compared later, it should accomplish simultaneity criteria of variability between the involved bands. Thus, \citet{Vil04} have considered time intervals lasting around 10\,min for photometric sequences between bands $V$ and $I$ in blazars, while \citet{Pap04} and \citet{Hu06} have calculated around 20 and 11\,min lags between bands $B$ and $I$ for AGNs and blazars, respectively. The difficulty of detection for even large microvariability events lasting less than 10 min has also been shown in the results of these simulations, no matter what the statistical methodology has been used (see for example Fig.~\ref{fig:det-vs-dur}). Thus, 10\,min is a safe time interval to bin data sharing similar flux characteristics, statistically indistinguishable from the noise, and will be appropriate for many studies of quasar microvariability. However, for the ANOVA continuous monitoring strategy, it is still possible to improve the test power by trying out different bin sizes \emph{after} the observations have already been made (\see \S\ref{sec:improving}).

    The choice for 1\,min exposures has been justified in \S\ref{sec:simulations}. Thus, a group of around 5 such exposures, lasting less than 10 min accounting for CCD read-out, is a reasonable methodological choice for the ANOVA with 30\,min lags observational strategy, for which the bin size of the groups is set \emph{before} the observations and cannot be changed after. For a larger telescope, the exposures might be shorter, but the total number of observations in each group is still limited by the read-out dead time. Besides, the gain in the power of the test would be relatively small for a number of exposures larger than 5.

    In the rest of this section only the \chisq test, the One--Way ANOVA with and without time lags between group observations, and the \Cstat\ will be discussed. The results of the simulations presented in \S\ref{sec:results} for the \chisq test using the standard deviation of the comparison star, and the $F$ test for sampled variances, are enough descriptive to show the possible risks of relying on these procedures.

    \subsection{Tests comparison} \label{sec:comparison}

    It is expected that the probability of detecting a change in the brightness of a source depends on both the amplitude of the variation and its duration. The range of amplitudes is the same for both the Gaussian peak and the linear simulations, but the duration of the Gaussian peak variations is restricted to a FWHM of 60\,min or less, while the duration of the linear variations spreads over the 5\,h of monitoring. Therefore it is not surprising then that linear variations are detected more easily than the shorter Gaussian peaks considered in the simulations (\see Table~\ref{tab:results}). The effect of the duration of the Gaussian peak variations on the number of detections is shown in Fig.~\ref{fig:det-vs-dur} for \chisq test, One--Way ANOVA, and One--Way ANOVA with 30\,min lags.

\begin{figure*}
\begin{center}
 \ifthenelse{\boolean{@twocolumn}}{\epsscale{2}}{\epsscale{1}}
 \plottwo{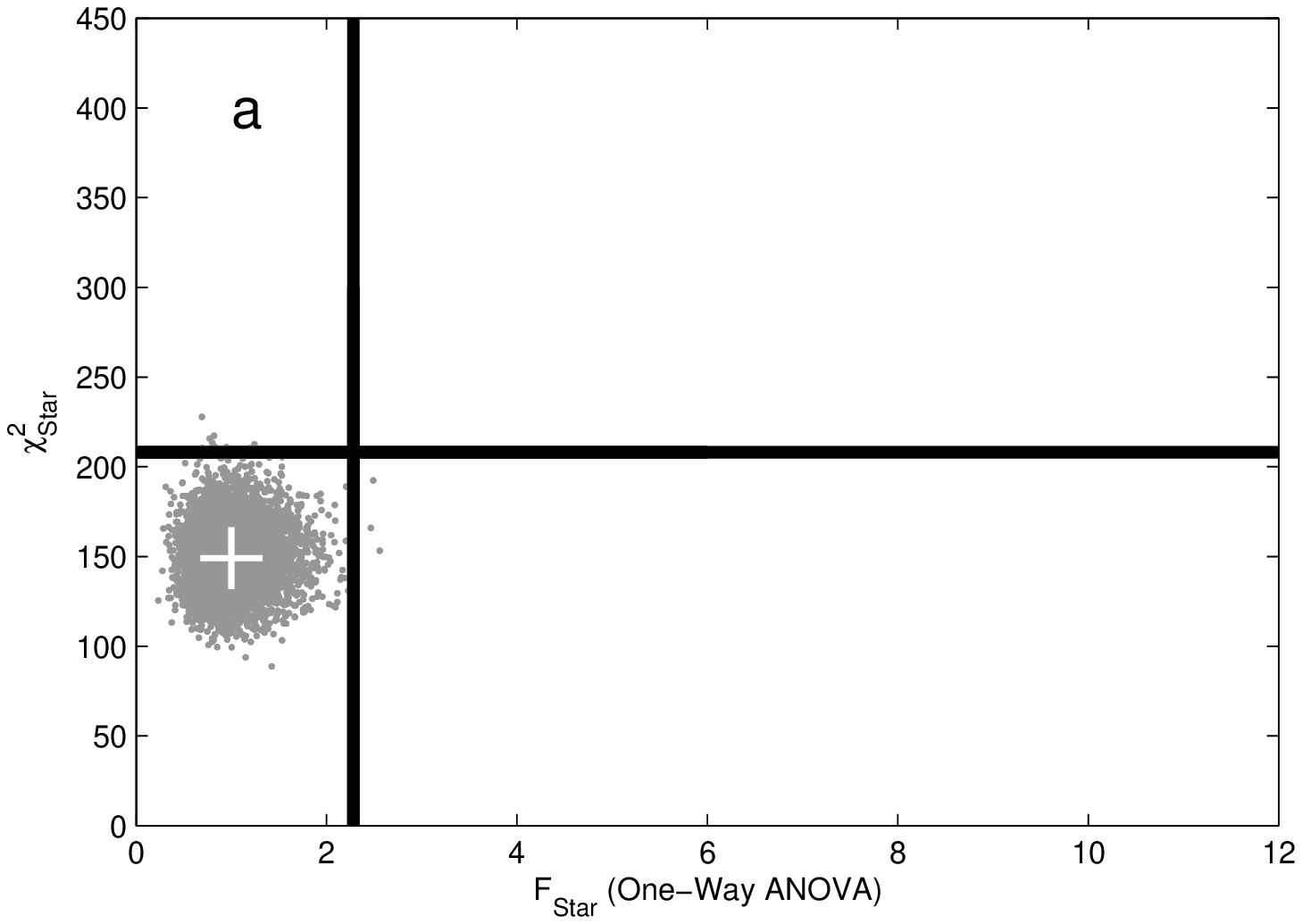}{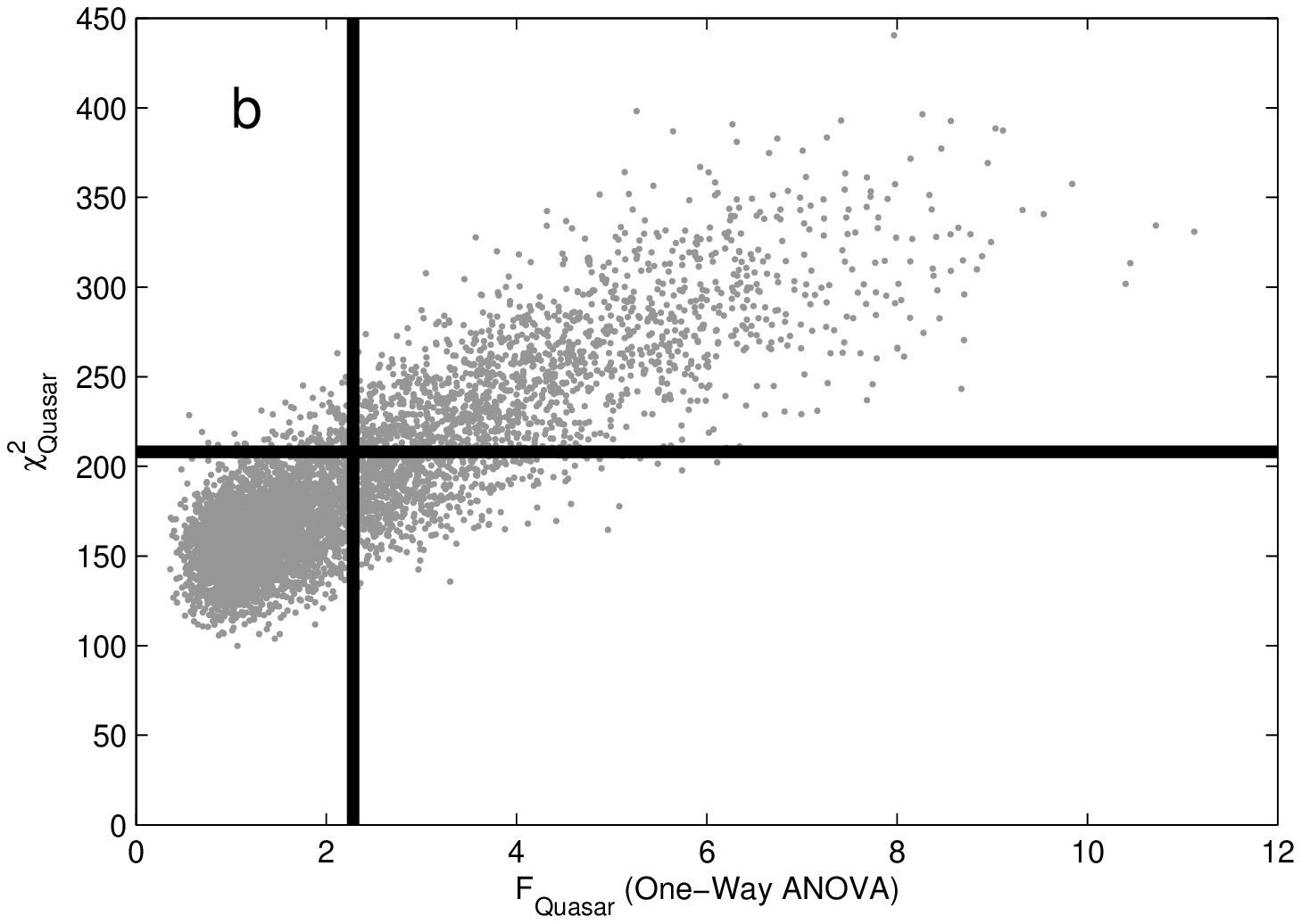}
\end{center}
\caption{Comparison between the \chisq and the One--Way ANOVA $F$ statistics for (a) reference star and (b) quasar with gaussian variations. Thick lines show the critical values for each statistics. The white cross in (a) marks the expected value of both statistics for a non-variable reference star. The distribution of data points in (b) corresponds to 3682 cases where neither \chisq nor ANOVA detect variations (lower left corner), 131 cases where \chisq detects variations and ANOVA does not (upper left corner), 704 cases where ANOVA detects variations and \chisq does not (lower right corner), and 1483 cases where both tests detect variations (upper right corner).} \label{fig:FvsChi2}
\end{figure*}

    Fig.~\ref{fig:FvsChi2}a compares the distribution of the \chisq and the \mbox{ANOVA $F$} statistics for the non-variable comparison star. This plot yields a circular shaped cloud of points (allowing a certain amount of distortion since the use of different scales), as expected for non-biased statistics. The points are centered at coordinates (1,149), as the expected values for \mbox{ANOVA $F$} and \chisq with 149 degrees of freedom, when the null hypothesis is true, are 1 and 149, respectively. Fig.~\ref{fig:FvsChi2}b shows a similar plot for the Gaussian peak variation quasar statistics. The clear correlation between both statistics indicates that they are measuring the same variable phenomenon. Note that the maximum range of the \chisq statistics is about 2 times its critical value, while the \mbox{ANOVA $F$} statistics spreads out approximately 5 or 6 times from its respective critical value. This is consequence of the different powers of the tests under the conditions of these simulations. The same arguments apply in the case of the linear variations. In this case, the difference in power between the ANOVA and the \chisq tests can be illustrated from the results for the quasar lightcurve shown in Fig.~\ref{fig:lc_linear} where the ANOVA $F$ statistics is larger than the critical value set to detect variations ($F = 2.8 > 2.3 = F^{(0.001)}_{29,120}$), while \chisq is below ($\chi^2 = 164 < 208 = \chi^2_{0.001,149}$).

\begin{figure}
\begin{center}
 \ifthenelse{\boolean{@twocolumn}}{\epsscale{0.75}}{\epsscale{0.45}}
 \plotone{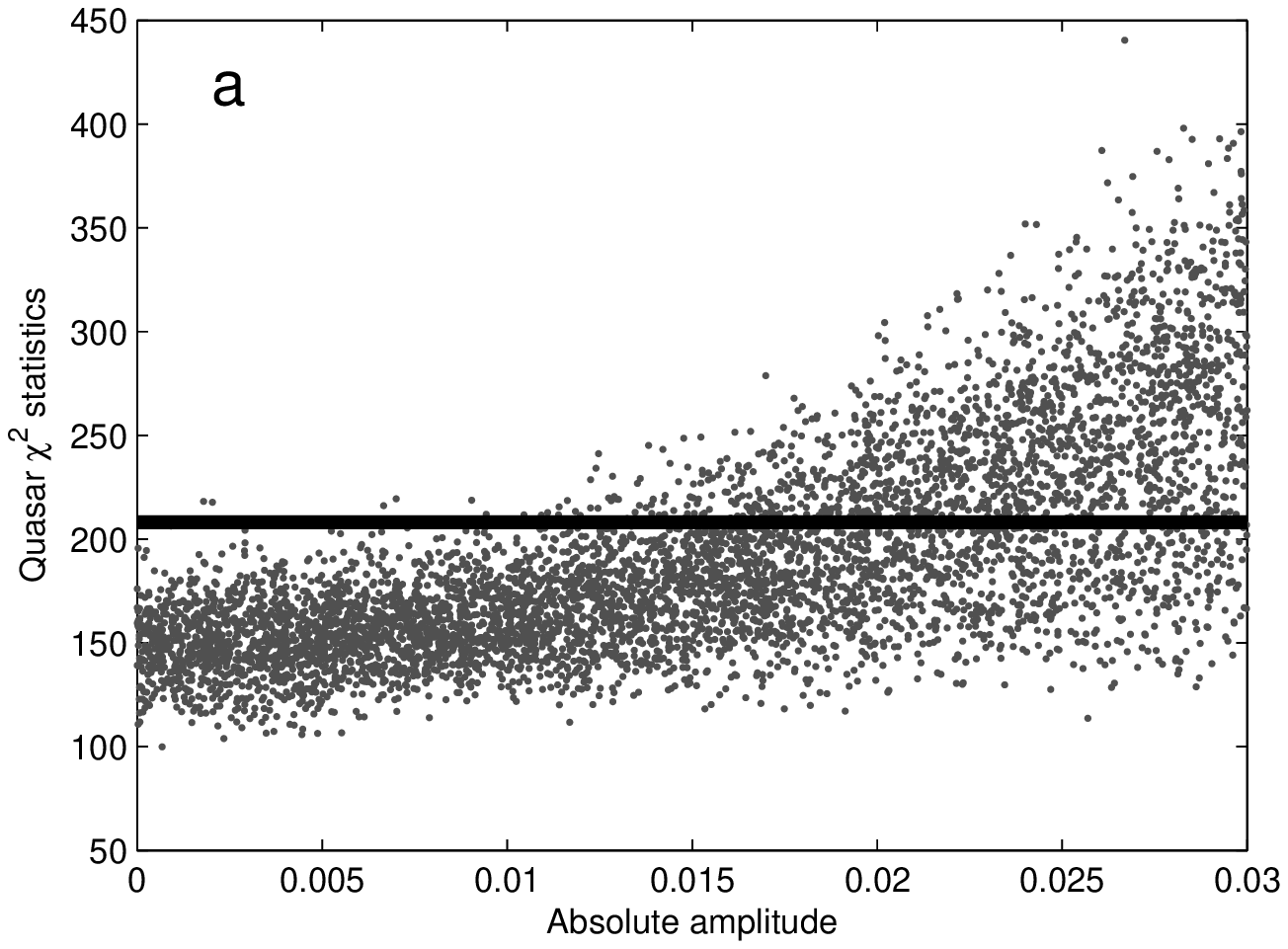}
 \vspace{-5pt}
 \plotone{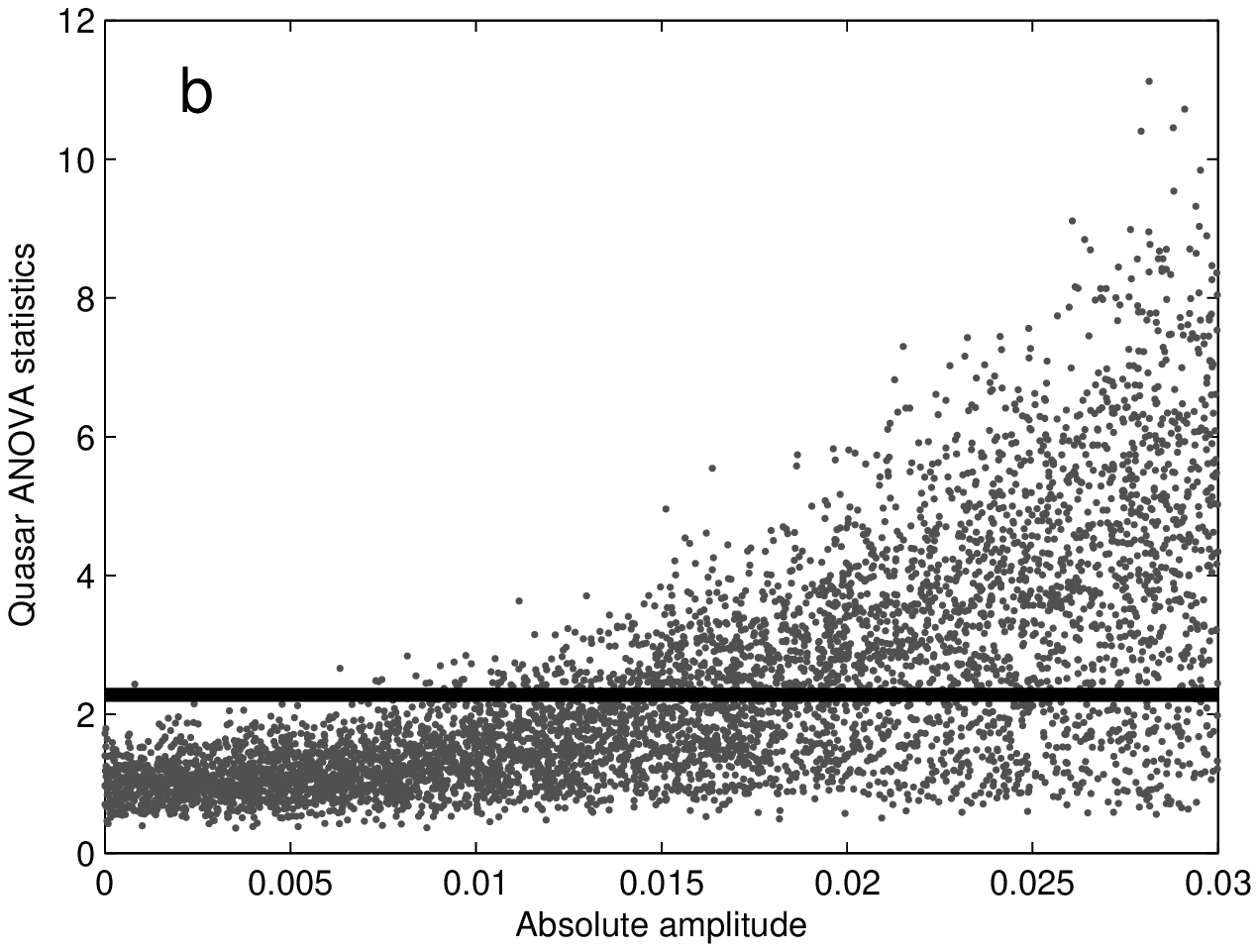}
 \vspace{-5pt}
 \plotone{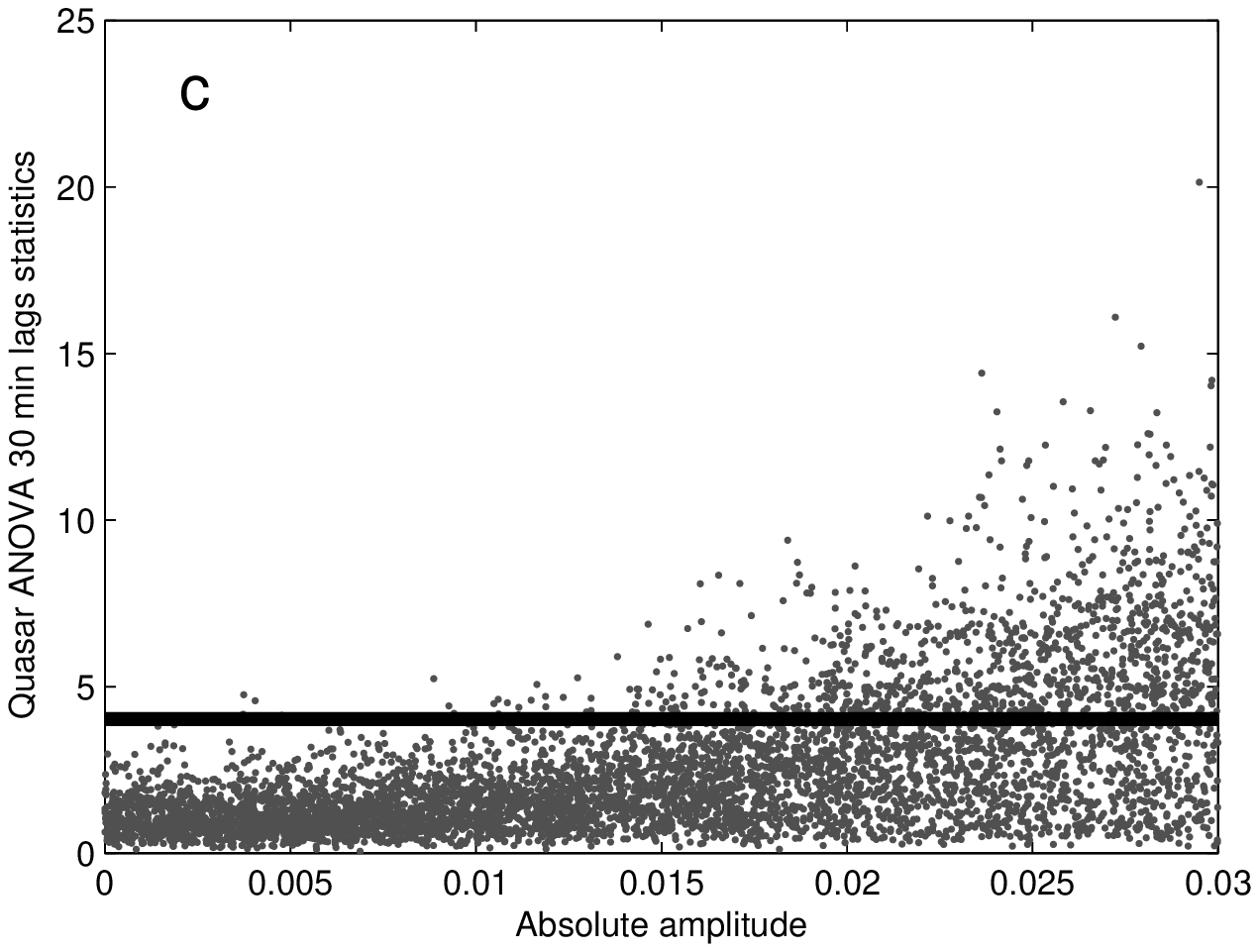}
 \vspace{-5pt}
 \plotone{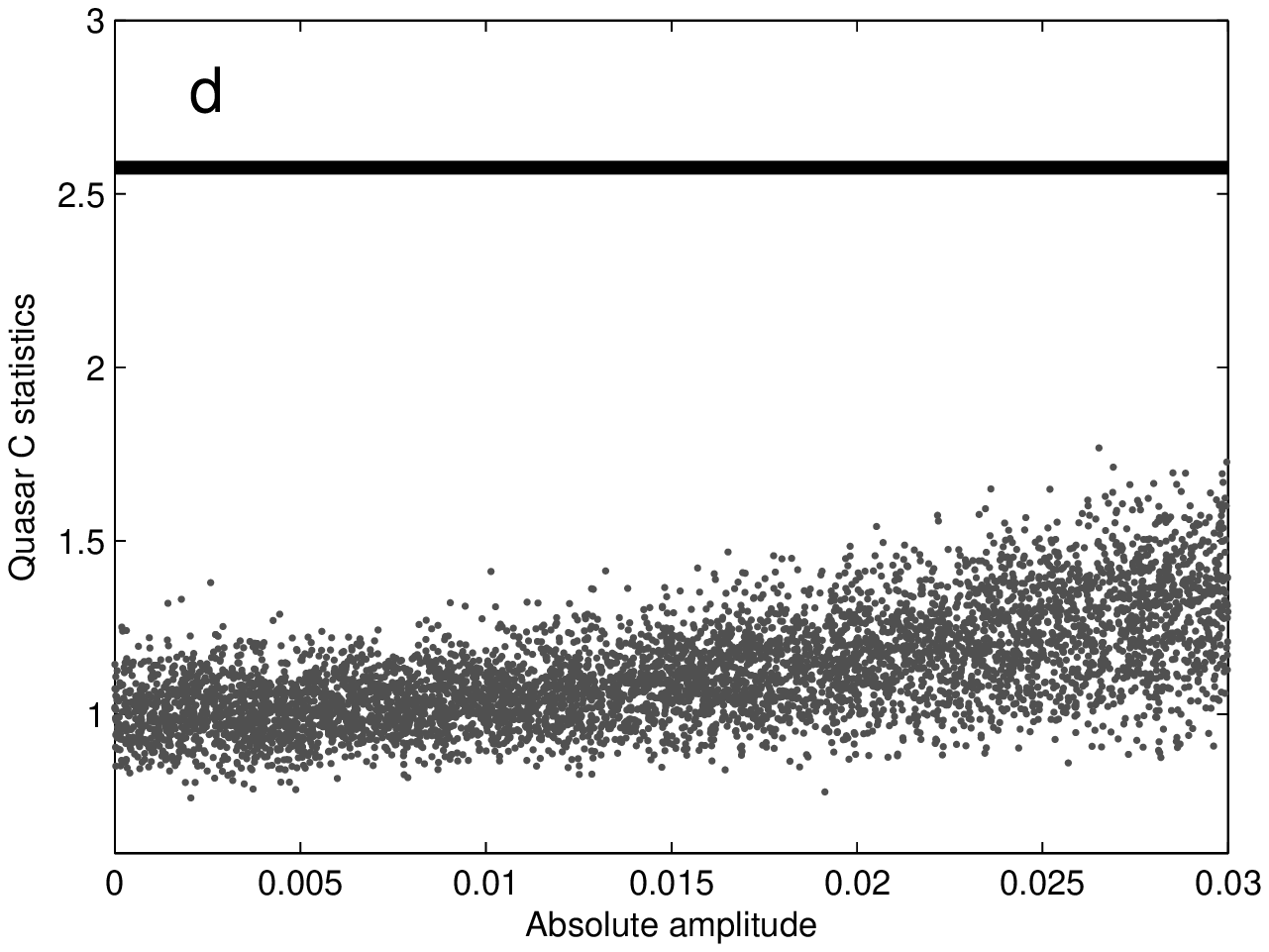}
\end{center}
 \caption{Distribution of statistics against absolute amplitudes of the Gaussian peak variations and critical values; (a) Points indicate the values of the \chisq statistics with $\alpha_1 = 0.001$, $\nu = 149$, and the solid line the critical value for the one--sided test $\chi^2_{\nu,\alpha_1} = 208$; (b) Idem for the ANOVA statistics with $\alpha_1 = 0.001$, $\nu_1 = 29$, $\nu_2 = 120$ and $F_{(\nu_1,\nu_2)}^{(\alpha_1)} = 2.28$; (c) Idem for the ANOVA statistics with 30\,min lags, $\alpha_1 = 0.001$, $\nu_1 = 9$, $\nu_2 = 40$ and $F_{(\nu_1,\nu_2)}^{(\alpha_1)} = 4.02$; (d) Idem for the \Cstat\ with $\alpha_2 = 0.01$ and its critical value for the two-sided normal test $z_{\alpha_2} = 2.576$.} \label{fig:stat-vs-ampl}
\end{figure}

    Fig.~\ref{fig:stat-vs-ampl} shows the distributions of (a) \chisq statistics, (b) \mbox{ANOVA $F$}, (c) \mbox{ANOVA $F$} with 30\,min lags between group observations and (d) \Cstat, against absolute amplitude of the variations for temporal Gaussian peak variations, along with the critical values for each test (indicated by thick horizonal lines). It is clear that for (a), (b) and (c) the number of detections (points above the lines indicating critical values) increases with the variability amplitude. Remember that ANOVA, \chisq and \Cstat\ were calculated using the same data set, while for One--Way ANOVA with 30\,min lags the data set is resampled to one third of the simulated observations. From Fig.~\ref{fig:stat-vs-ampl}d, it is obvious that \Cstat\ is about a factor 2 below any detection even though the nominal significance level for this test ($\alpha = 1\%$) is less tight than for ANOVA and \chisq tests ($\alpha = 0.1\%$).

\begin{figure}
\begin{center}
 \ifthenelse{\boolean{@twocolumn}}{\epsscale{0.75}}{\epsscale{0.45}}
 \plotone{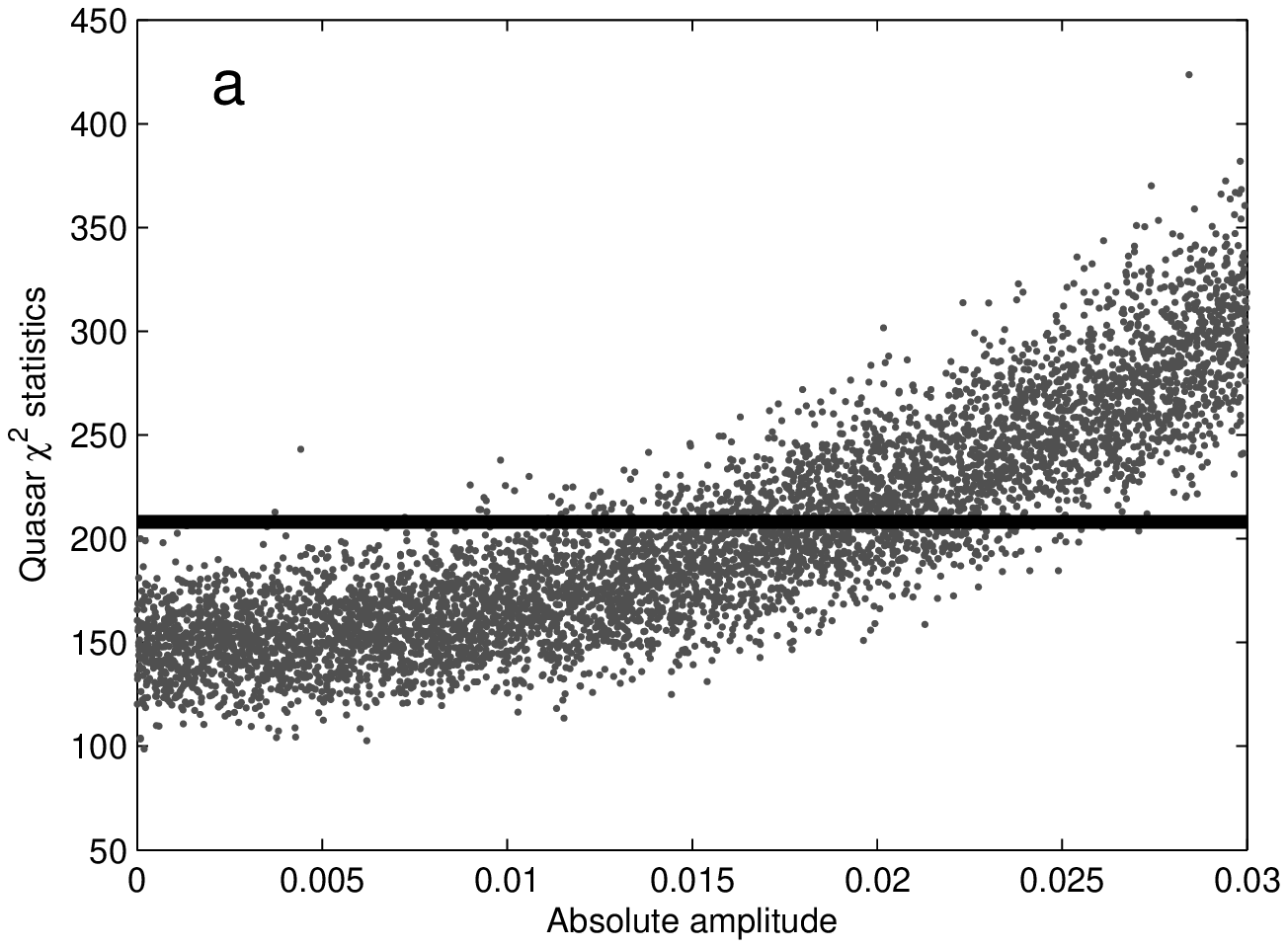}
 \vspace{-5pt}
 \plotone{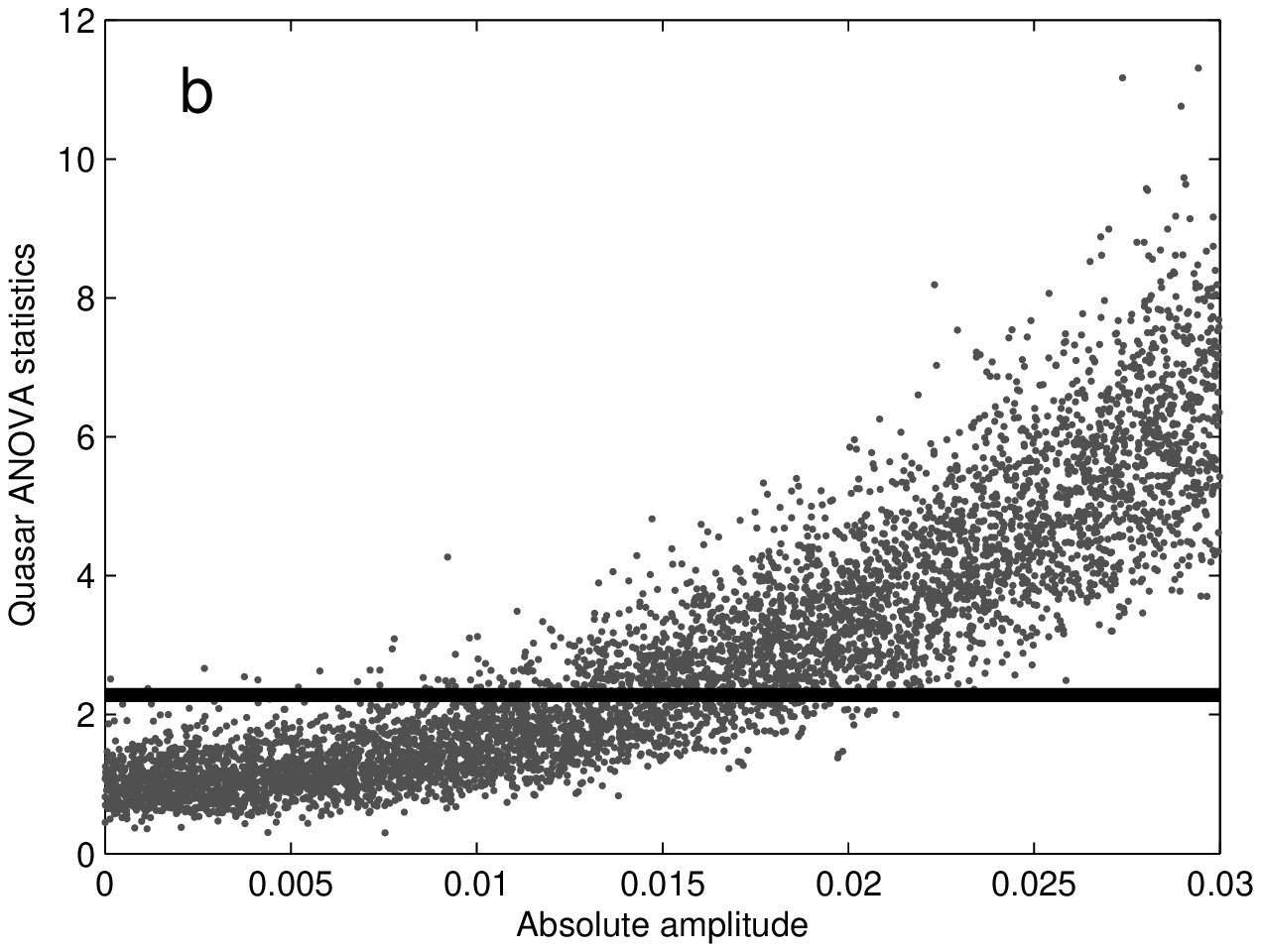}
 \vspace{-5pt}
 \plotone{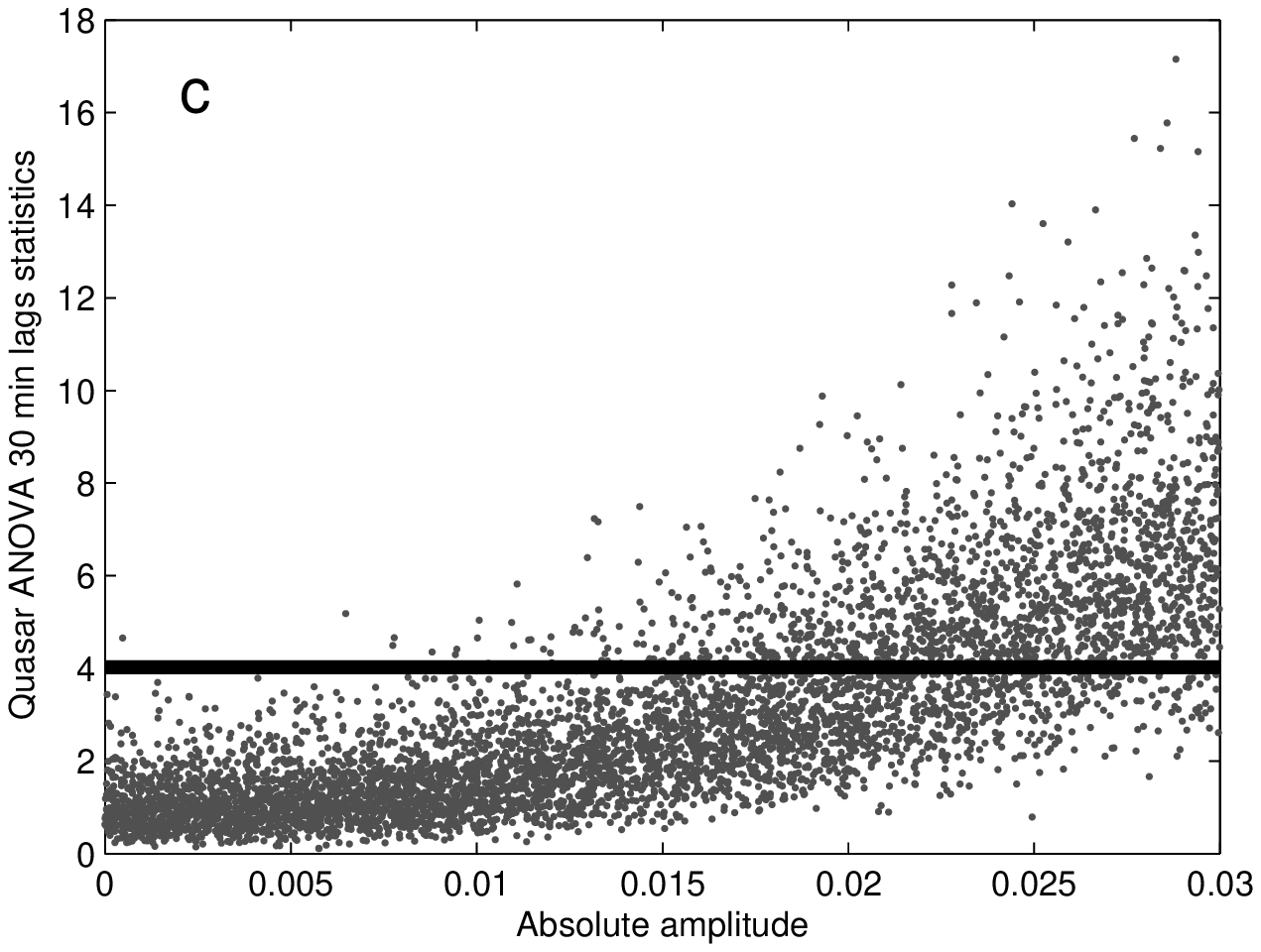}
 \vspace{-5pt}
 \plotone{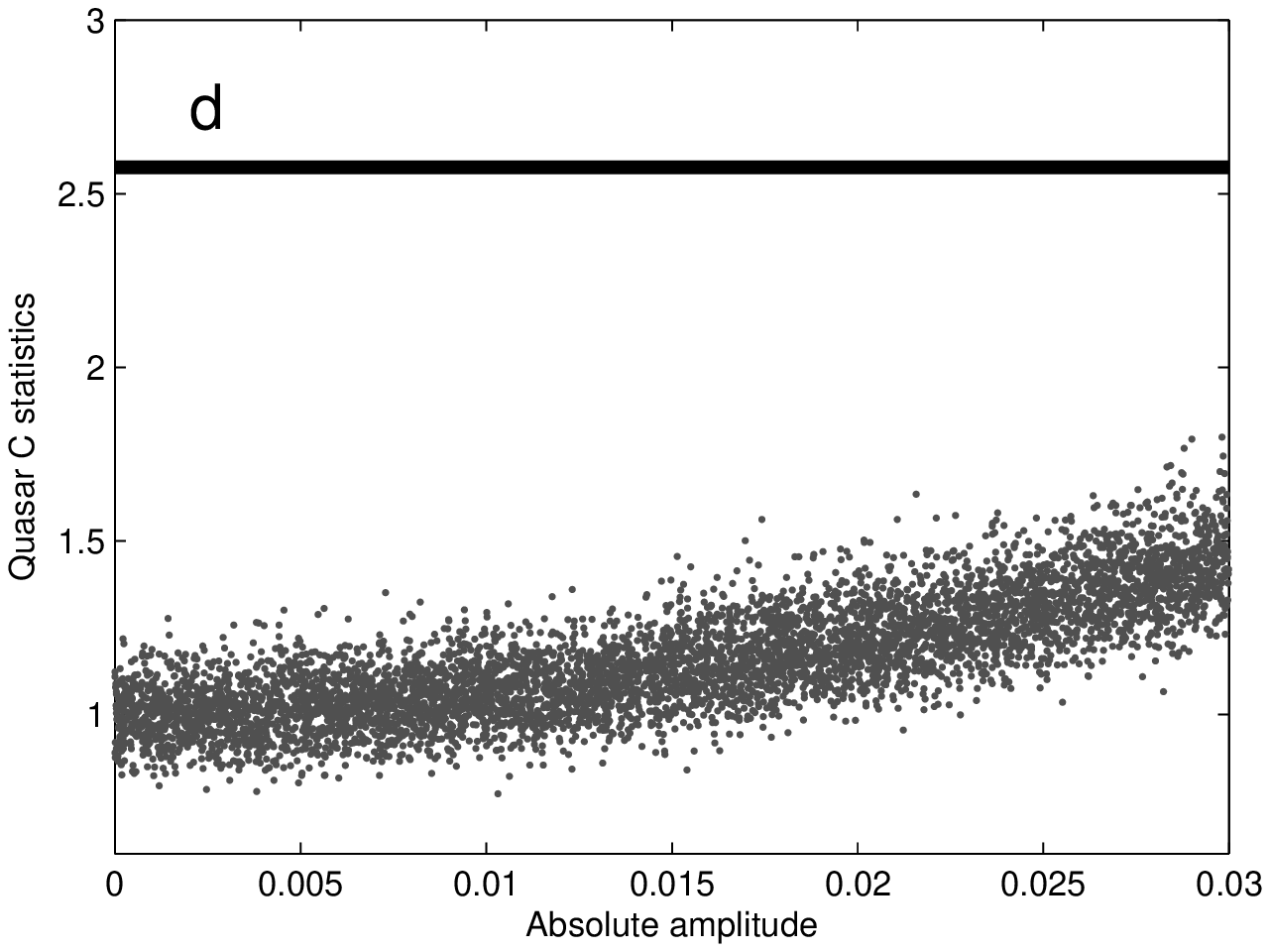}
\end{center}
 \caption{Distribution of statistics against absolute amplitudes of the linear variations and critical values; (a) Points indicate the values of the \chisq statistics with $\alpha_1 = 0.001$, $\nu = 149$, and the solid line the critical value for the one--sided test $\chi^2_{\nu,\alpha_1} = 208$; (b) Idem for the ANOVA statistics with $\alpha_1 = 0.001$, $\nu_1 = 29$, $\nu_2 = 120$ and $F_{(\nu_1,\nu_2)}^{(\alpha_1)} = 2.28$; (c) Idem for the ANOVA statistics with 30\,min lags, $\alpha_1 = 0.001$, $\nu_1 = 9$, $\nu_2 = 40$ and $F_{(\nu_1,\nu_2)}^{(\alpha_1)} = 4.02$; (d) Idem for the \Cstat\ with $\alpha_2 = 0.01$ and its critical value for the two-sided normal test $z_{\alpha_2} = 2.576$.} \label{fig:stat-vs-ampl-lin}
\end{figure}

    The results for the linear variations shown in Fig.~\ref{fig:stat-vs-ampl-lin} are similar to Gaussian peak variations, but the statistics are less scattered because there is no effect of the length of the variation. For a significance level of $\alpha = 0.1\%$ and the conditions of the simulations, all the linear variations with amplitudes $\gtrsim 0.027$ are detected by the \chisq test. On the other hand, One--Way ANOVA detects all the variations with amplitudes $\gtrsim 0.022$. In comparison, One--Way ANOVA with 30\,min lags detects around 90\% variations for amplitudes near 0.03. As in the Gaussian peak case, \Cstat\ is again a factor 2 below any detection at the significance level of $\alpha = 1\%$.

\begin{figure}
 \ifthenelse{\boolean{@twocolumn}}{\epsscale{0.8}}{\epsscale{0.5}}
  \plotone{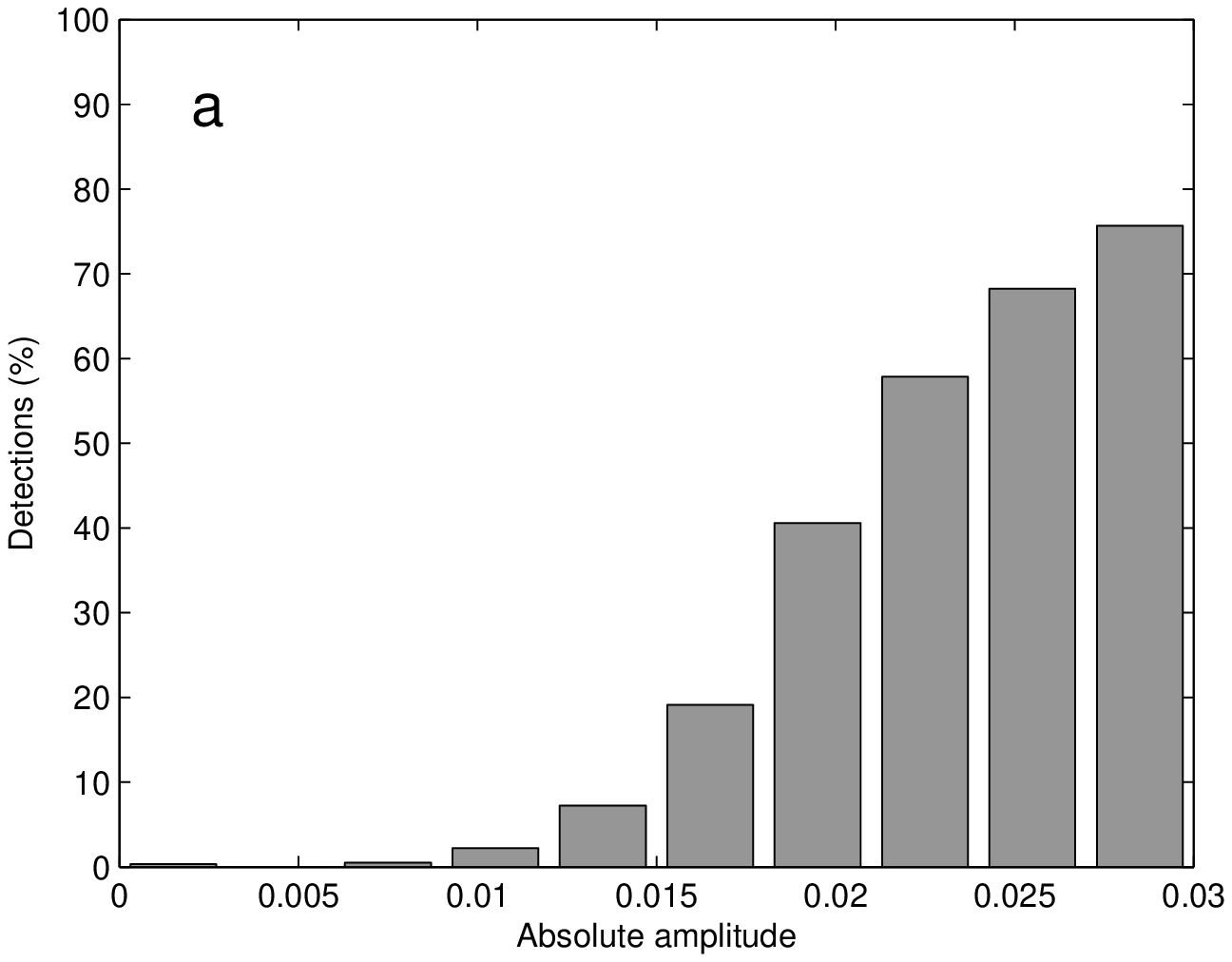}
  \plotone{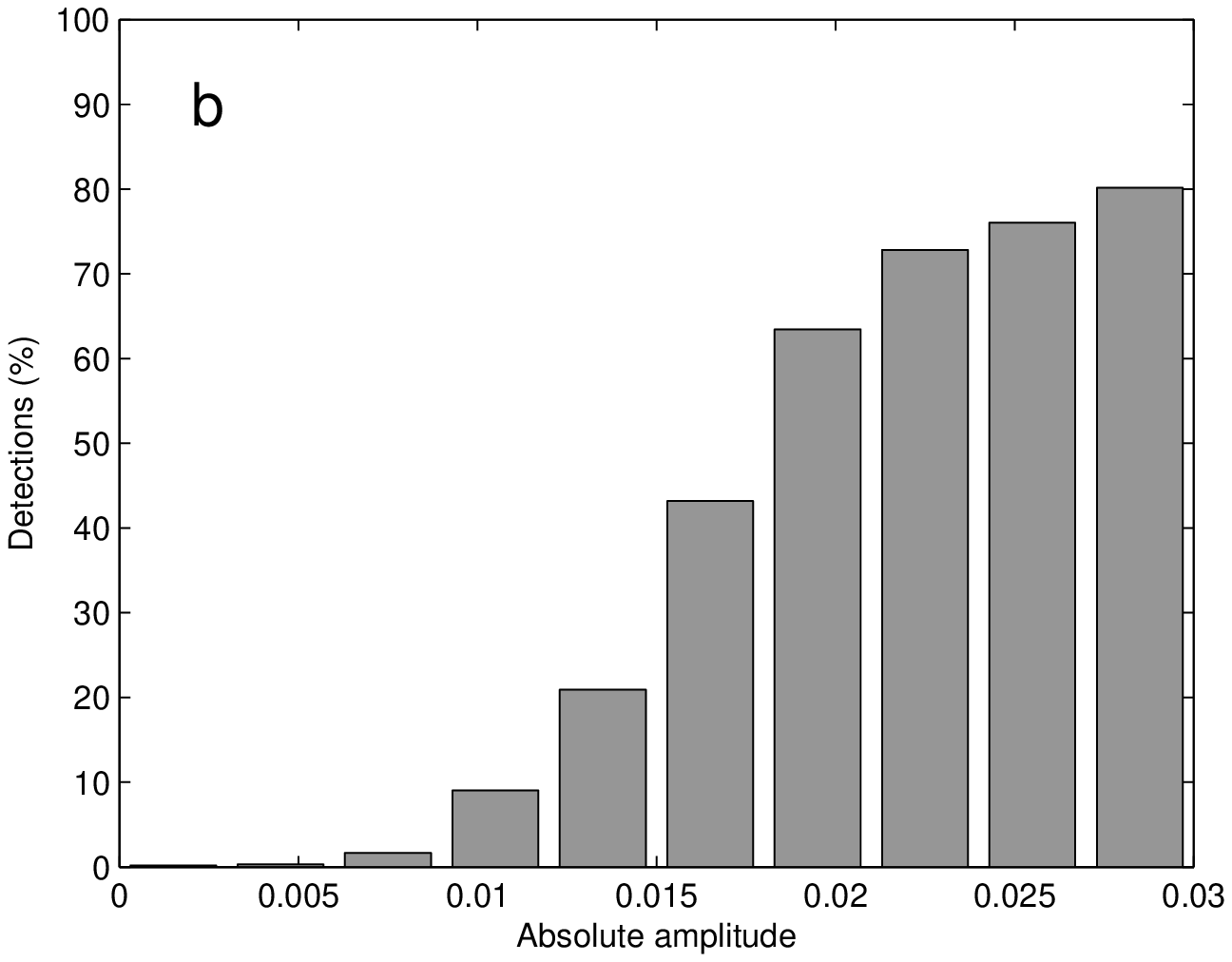}
  \plotone{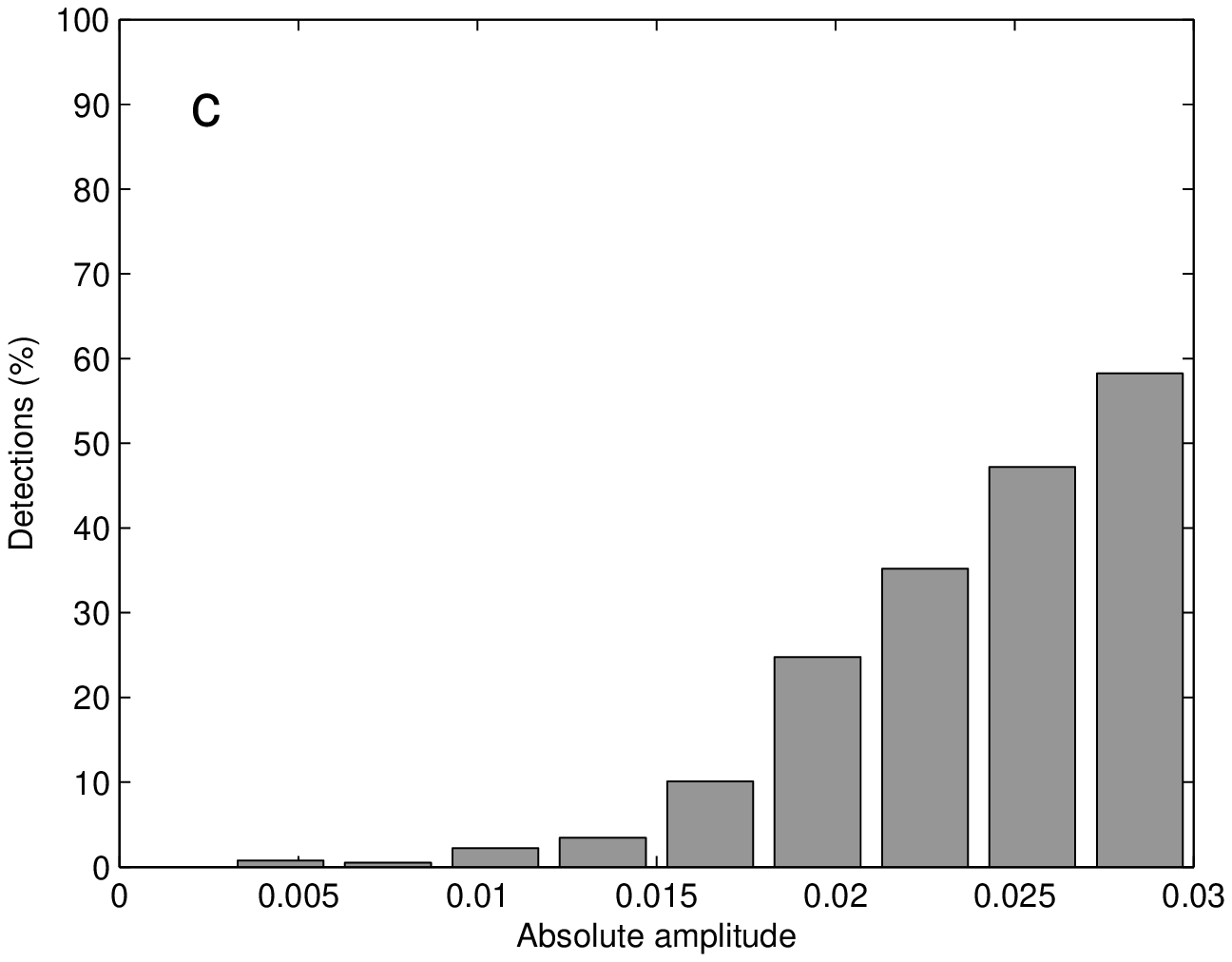}
  \caption{Percentage of detections per amplitude of Gaussian peak variation. (a) \chisq statistics; (b) ANOVA; (c) ANOVA with 30\,min lags.}\label{fig:det-per}
\end{figure}

\begin{figure}
 \ifthenelse{\boolean{@twocolumn}}{\epsscale{0.8}}{\epsscale{0.5}}
  \plotone{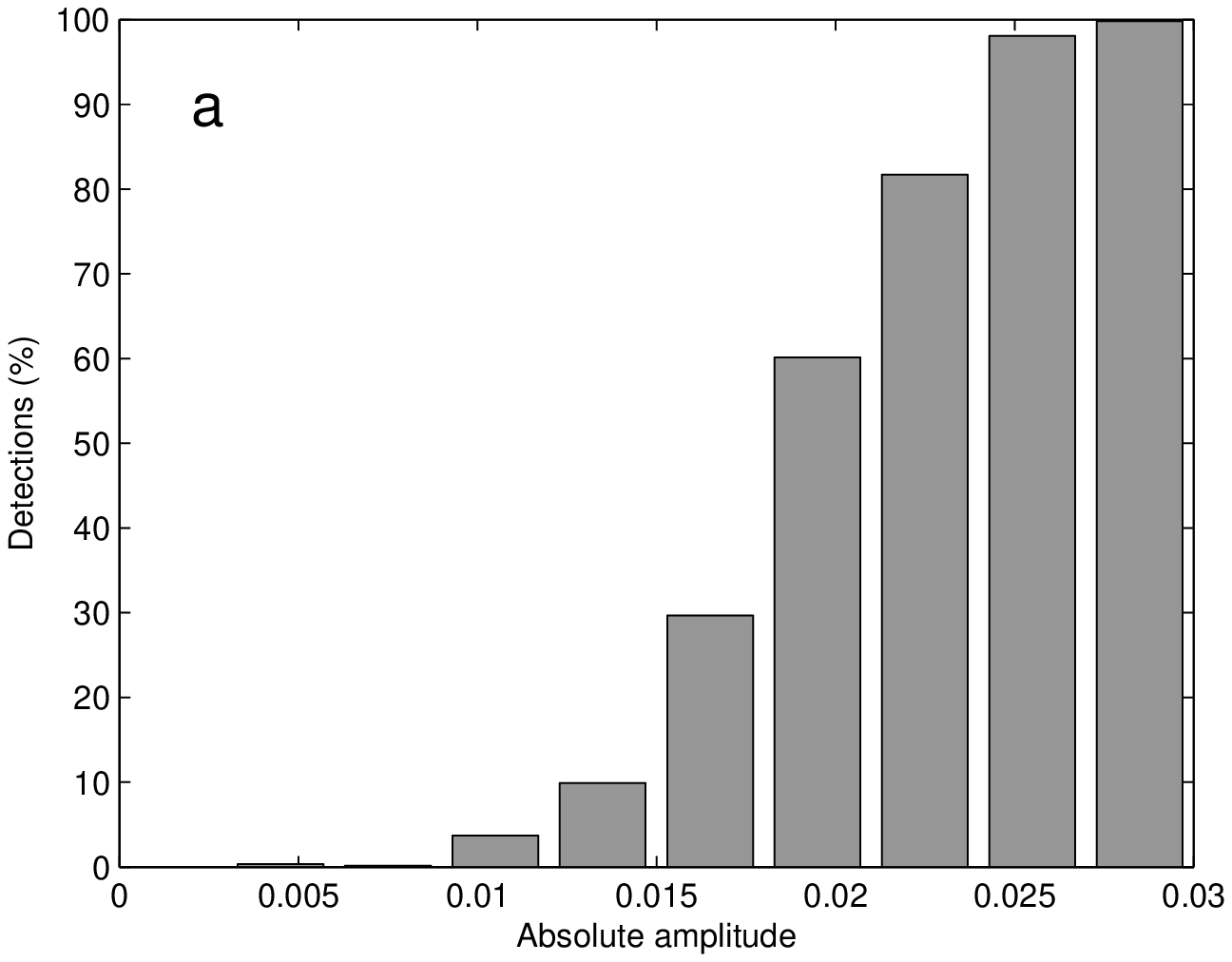}
  \plotone{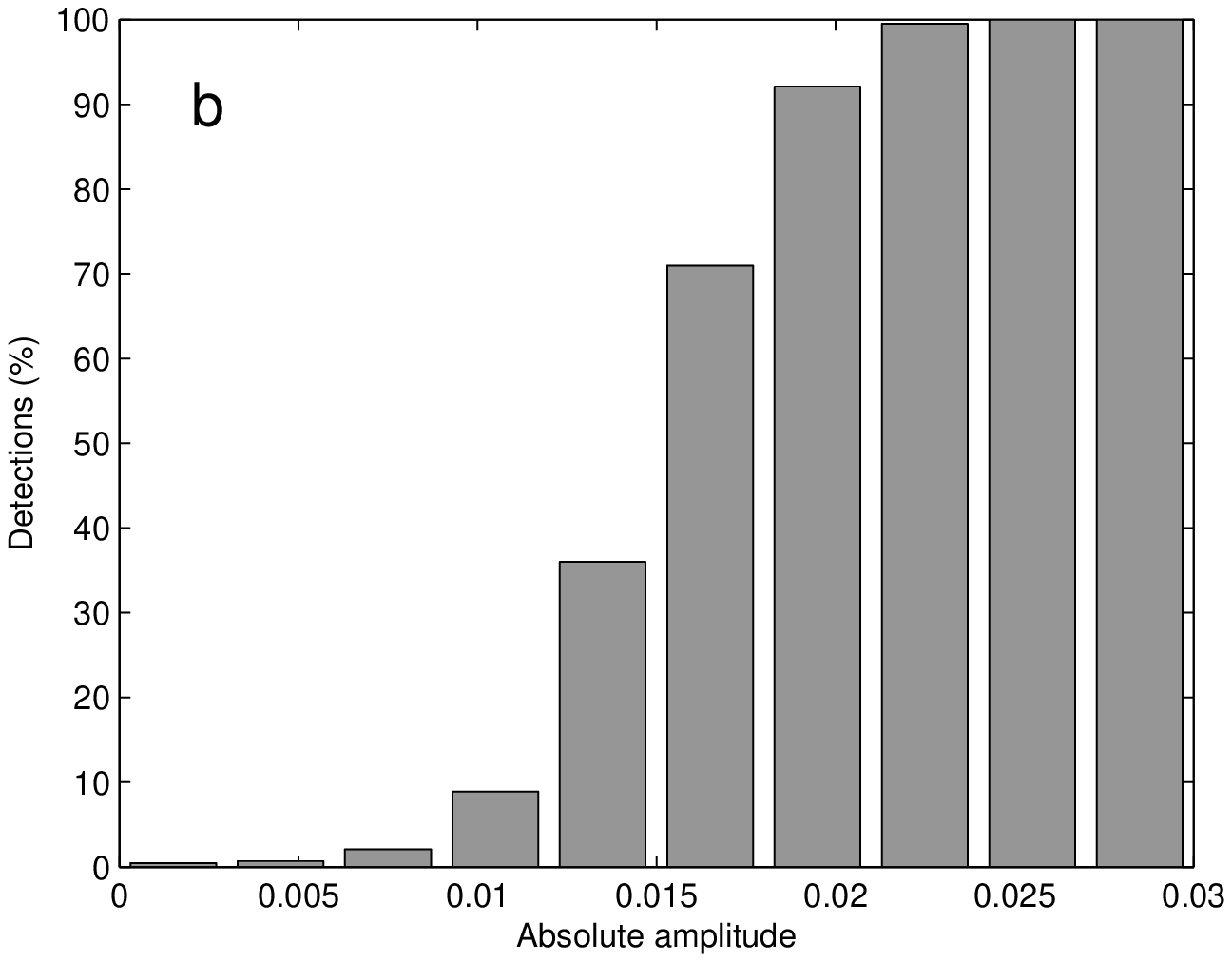}
  \plotone{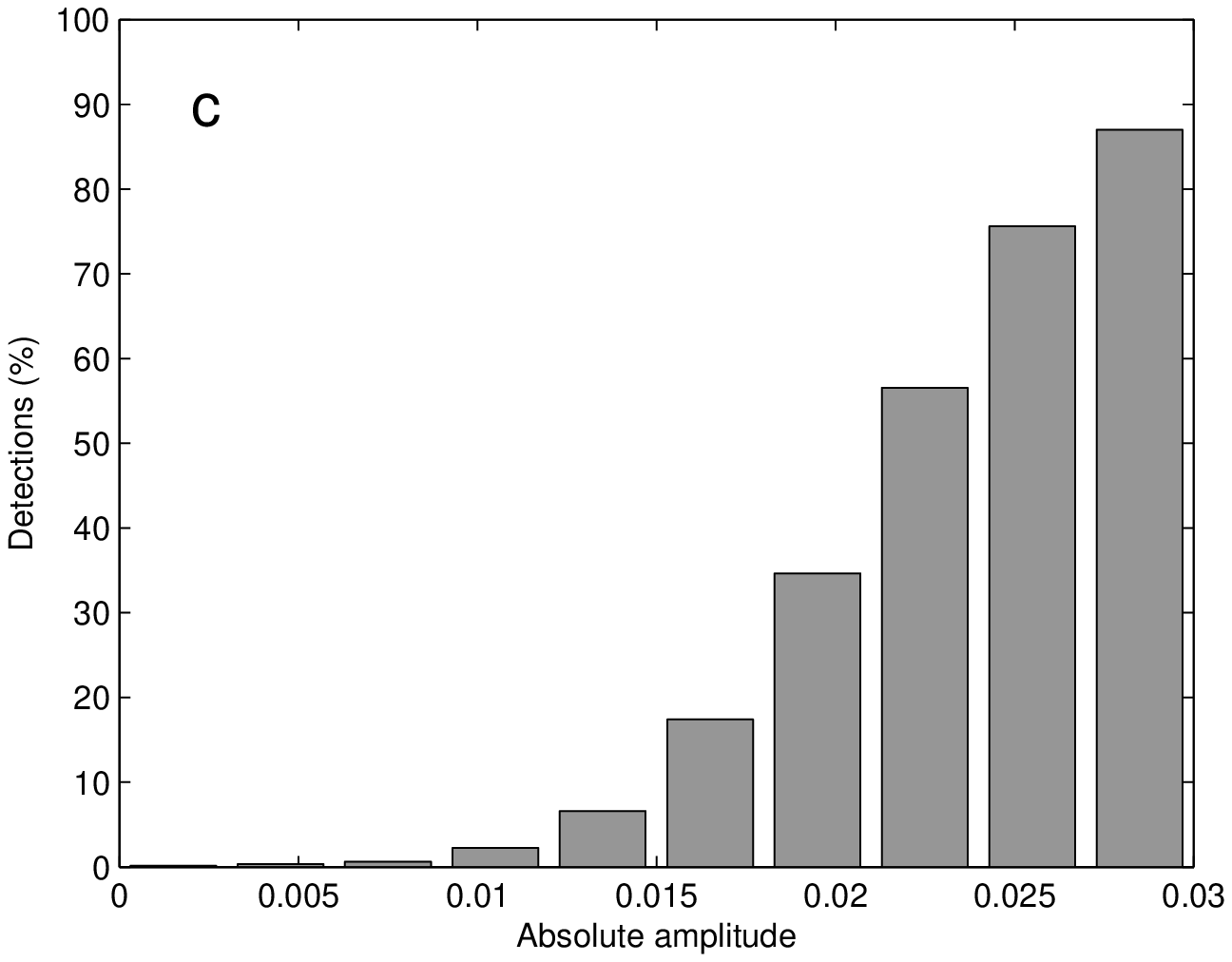}
  \caption{Percentage of detections per amplitude of linear variation. (a) \chisq statistics; (b) ANOVA; (c) ANOVA with 30\,min lags.}\label{fig:det-per-lin}
\end{figure}

\begin{figure}[t]
  \epsscale{0.95}
  \plotone{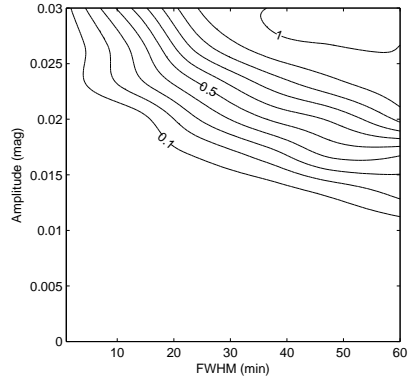}
  \caption{An example of the double dependence of the Gaussian peak detections on the amplitude and the temporal length of the variation. The smoothed \chisq statistics of the simulated data is shown as a contour plot of the probability of detection with FWHM of the variations in minutes in the X axis, and the amplitude in magnitudes in the Y axis. Plots for One--Way ANOVA statistics would show a similar aspect.}\label{fig:3D}
\end{figure}

    Percentages of detections per amplitude range are shown for the Gaussian peak variations in Fig.~\ref{fig:det-per}, and for the linear variations in Fig.~\ref{fig:det-per-lin}. Percentages for the Type\,II errors per amplitude range can be easily derived from these figures as the subtraction of the percentages of detections from one hundred. In the case of the linear variations considered in these simulations, the dependence of the number of detections with the amplitude of the variations is straightforward. But in the case of the Gaussian peak, the double dependence on the amplitude and the temporal length of the variation makes the relationship less evident. After smoothing, this double dependence can be shown as a contour plot of the probability of detection as a function of the amplitude and duration of the microvariability event, as shown in Fig.~\ref{fig:3D}. Note that the probability of detection increases with both the amplitude and the duration of the variation.

%
%

    The bulk of all these results attests that both the \chisq test and One--Way ANOVA are robust methodologies to study variability in the lightcurves of quasars. However, the \chisq test relies on an accurate theoretical estimation of the data error {\it for each single data point}. As commented in \S\ref{sec:chi2-test} this is not usually the case, and a number of \emph{ad hoc} factor corrections have been used to compensate the notorious lack of agreement between the IRAF estimated photometric errors and the actual dispersion of the data \citep[\eg][]{Gop03, Sta04, Bac05}.
    But the large increment in the number of Type\,I errors obtained using a sample standard deviation drawn from the comparison star instead of the actual error (Table~\ref{tab:results}), as well as arguments given above about the loss of power of the test when the data is binned, warns against any simple approach to dodge this problem. However, the error estimation issue is offset by the internal error estimation of the ANOVA tests. It is reasonably to assume for the \chisq test that an accurate estimate of the actual errors for each data point might be achieved by measuring the data dispersion for a large number of foreground stars. However, we have already discussed above that the dispersion measurements such as variances and standard deviations are lousy constrained. Let us investigate the possible gain in accuracy by using a set of comparison stars.

    Actually, the significance level of the \chisq test when the true photometric error of the data is unknown, but estimated from one or more comparison stars is very easy to calculate: it corresponds to the $F$ statistics. In our case, each lightcurve comprises 150 observations ($\nu = 149$ degrees of freedom) and the nominal significance level of the test is $\alpha = 0.1\%$, which corresponds to a critical $\chi^2_{0.001,149} = 208$. We divide this value by $\nu$ to obtain the \emph{reduced} \chisq critical value $\chi^2_r = 1.40$, and calculate the significance level $\alpha$ for the $F$ statistics with $\nu_1 = 149$ and $\nu_2 = 149 \times N_*$, for $F^{(\alpha)}_{149,149 N_*} = 1.40$. Some of these calculations are presented in Table~\ref{tab:combStars}. Column (1) indicates the number of stars $N_*$ used to estimate the error; column (2) is the actual significance level $\alpha$ for the test; and column (3) is the relative error in the variance estimate.

    Note the value for a single comparison star ($N_* = 1$); the actual significance level is 2.12\% rather than 0.1\%, therefore, for 6000 lightcurve simulations this test should produce $0.0212 \times 6000 \approx 127$ Type\,I errors, which agrees with the result of the simulations for the \chisq test performed using the standard deviation of the comparison star (\see the second item in Table~\ref{tab:results}). Other interesting results reported in Table~\ref{tab:combStars} are that combining the data for 10 stars, the actual significance level is still almost twice the nominal value, and that it is necessary to combine up to 30 or 40 stars to obtain a significance level accurate up to 20\%.

\begin{table}[t]
 \centering
 \begin{minipage}{0.85\linewidth}
  \caption{Effect of the number of \newline comparison stars on the test accuracy. \label{tab:combStars}}
  \begin{tabular}{rrr}
\tableline\tableline
    \multicolumn{1}{c}{Number}  \\
    \multicolumn{1}{c}{of stars} & \multicolumn{1}{c}{$\alpha$} & $\mathrm{Err}(s^2)/\sigma^2$ \\
\tableline
     1  &   0.0212  &   1.4142 \\
     5  &   0.0029  &   0.6325 \\
    10  &   0.0018  &   0.4472 \\
    20  &   0.0014  &   0.3162 \\
    30  &   0.0012  &   0.2582 \\
    40  &   0.0012  &   0.2236 \\
\tableline
\end{tabular}
\end{minipage}
\end{table}

    Sample variances $s^2$ are \chisq distributed random variables (equation~\ref{eq:var_dist}) and therefore it can be demonstrated easily that, in our case, the sample variance for each photometric data point obtained from measuring $N_*$ stars has also an associated variance (\ie variance of the variance) given by:
    \[
    \mathrm{Var}[s^2] = \sigma^4 \frac{2}{N_*-1},
    \]
    where $\sigma$ is the true photometric error, and $N_*-1$ the degrees of freedom. Thus the error of the measured variance is expressed by the square root of the previous equation: $\mathrm{Err}[s^2] = \sigma^2 \sqrt{2/(N_*-1)}$. For observations analogous to the simulations reported in this paper, $\sigma \approx 0.01$\,mag. The relative error for the variance estimate ($\mathrm{Err}[s^2] / \sigma^2$) obtained from $N_*$ star is shown in the third column of Table~\ref{tab:combStars}. Even for $N_* = 40$ the variance estimate has an accuracy worse than 20\%, which implies that the photometric error estimate is inaccurate by approximately 50\% ($s = 0.010 \pm 0.005$). But even in the case that it were possible to observe tens of stars simultaneously, attempting to meet the controlled conditions of the simulations where the errors are completely known, ANOVA's statistical power performs better than the \chisq test to find tiny variations in the quasar simulated differential lightcurves. Although the actual differences in power may vary depending on the test implementation and lightcurve characteristics, or even the actual observational methodology, the effort of combining the lightcurves of tens of foreground stars may be irrelevant in most cases.

    Another concern with measuring errors using a large sample of suitable foreground stars is that it is not always possible. In fact, the number density of bright stars around quasars that can be used for differential photometry studies is small. From a photometric point of view, quasars are blue color objects. For the most reliable differential photometry, at least in the most blue optical bands, quasars should be compared preferable with either nearby white dwarf stars or Main Sequence bluish stars to avoid color effects that may arise at different air-masses. Besides, quasars are observed at high galactic latitudes and thus, unless the telescope field is large enough, the number of foreground stars around quasars is usually scarce. Moreover, many of them may be old, low luminosity stars hanging around the thick galactic disk or roaming through the galactic halo. Then, many of the most brightest foreground stars will be luminous  Pop\,II red giants and evolved, fast period variable stars. Although the effect of observing objects with moderate color differences at low air-masses  is negligible in broad band studies, red giant stars should be avoided in differential optical photometry of quasars. Fast period variables are of course unsuitable as reference and comparison stars. Therefore, around most quasars there are only a few useful nearby, not too red and non-variable stars bright enough to be used for comparison purposes.

    Another interesting subject is the possibility of detecting very fast variations. It has been commented above that the statistical power of the tests is limited by both the duration of the microvariability event and its amplitude. For the observational parameters considered in these simulations, the effort to detect microvariations lasting less than $\approx$30\,min and with amplitudes of less than 0.02\,mag would be very inefficient. The relatively large number of spikes reported in microvariability literature \citep{Sag96, Die98, Gop00, Sta04} suggests that they may be a common phenomenon and worth of investigation using very fast and accurate photometry with large telescopes.

    The One--Way ANOVA with 30\,min lags procedure is analogous to the methodology used by \citet{Die98}. Simulations show that this procedure is also robust, but the smaller number of points in each data set (50 observations rather than 150) due to the discontinuity in the sampling affects the number of detections, which is about one half of the One--Way ANOVA continuous sampling. But in the real world, the final result will depend on the actual timescales of the microvariations. \citet{Sta04} report three kind of variations, namely small gradual variations lasting over several hours, time resolved microvariability on hour-like timescales, and single-point fluctuations (spikes). In this case, One--Way ANOVA with 30\,min lags will perform almost as well (within a 20\%) as the continuous sampling detecting the first two types of microvariations. Therefore, as this technique permits to monitor other target fields during the gaps between the end of a group of observations and the beginning of the next group, it turns out to be a very powerful exploratory methodology to detect microvariations. For example, in the One--Way ANOVA with 30\,min lags case each group of 5 observations lasts 10\,min (accounting for exposures and read-outs), and there is a gap of 20\,min until the first exposure of the next group begins. Depending on the telescope setup, it would be possible to switch to at least one, maybe two near different target fields, increasing the number of monitored objects by a factor two or three.


    The results for ANOVA and \chisq tests discussed above do not demonstrate that, under all circumstances, One--Way ANOVA tests perform better than a well implemented \chisq test. It is possible to sustain that similar results to those presented in this paper would have been obtained in the case, for example, of lightcurve shapes not considered here, such as sawtooth variations. However, there are many possibilities of test implementation and observational techniques. One of the requirements of ANOVA tests is the homogeneity of variances, that is, errors should be distributed equally in all the groups. ANOVA is robust against moderate violations of this requirement, and thus this is not a big concern when comparing data obtained with the same telescope and equipment, and during the same night under ordinary atmospheric conditions. However, combining data from a couple of telescopes with different characteristics may ruin the One--Way ANOVA's performance.

    \subsection{Remarks on the \Cstat\ results} \label{sec:C-remarks}


    Proves on the reliability of the \Cstat\ methodology have confirmed the severity of the problems pointed out in \S\ref{sec:C-test}. This methodology has been widely accepted as a standard test, and thus it is not surprising the confusion generated in the research of microvariability in quasars. Several researchers have used \Cstat\ to study messy samples of radio quiet and radio loud quasars altogether with blazars. For example, \citet{Rom99} include several BL\,Lac objects, accounting for nearly 70\% of their microvariability detections. In most cases, only blazar-like objects will show variations extreme enough to lie above the proposed \Cstat\ critical value. A detailed discussion about the relevance of a careful sample selection can be found in \citet{Ram09}.

    In the case of high accuracy photometry, as the observations presented in \citet{Rom99} ($\sigma_e \sim 0.001\,\rm{mag}$), maybe rare microvariability events could be detected by the \Cstat\ also in non blazar objects. For example, \citet{Gup05} detect microvariation events if the errors are about 0.005\,mag, but usually fail to detect if errors are 0.01\,mag or above. Note that this result is in accordance with those presented in \S\ref{sec:results} in the sense that for the accuracy of 0.01\,mag considered in the simulations, \Cstat\ is always a factor $\gtrsim2$ below its critical value.

    There is contradictory evidence that some RQQ may have a weak blazar component \citep{Cze08, Cha09} which might explain at least some reported microvariability detections. If this blazar component is present, it might account for flux fluctuations in RQQs above 0.05\,mag, and the extreme variations above 0.1\,mag reported for PG\,0026+129 \citep{Jan05} and US\,995 \citep{Die98} (however, note that large amplitude variations are not considered in the simulations presented in this paper).

    When a more appropriate methodology is used, the results converge in the sense that microvariability in radio quiet quasars is also a common phenomenon. Thus, \citet{Gup05} used the $F$ test and found that 2, probably 3 radio quiet quasars out of a small sample of 6 present microvariations (\ie between 30 and 50\%), but when they mix their results with those obtained by other groups that use \Cstat, the total numbers are 6, possibly 14 objects that vary out of 49 (\ie between 12 and 29\%). On the other hand, \citet{Sta05} separate BL\,Lacs in their sample and, using a combination of the \Cstat\ and the \emph{structure function} find that the Duty Cycle for radio loud and radio quiet quasars are not statistically different (18 and 22\% respectively). This result is also analogous to the value calculated by \citet{Sta04} for the Duty Cycle inferred from \citet{Die98} ($\sim$25\%).

\section{Conclusions} \label{sec:conclusions}

    In this paper, several statistical methodologies and their variants for microvariability studies in quasars have been tested. From these techniques, One--Way ANOVA and \chisq tests have shown to be the most robust statistics. However, some 'reasonable´ \chisq tests variants, particularly those that estimate errors from the lightcurves of comparison stars rather than rely on accurate physical models to characterize the data quality, are much less robust. These \chisq variants should be avoided, as they can be substituted easily by other robust tests, or at least they should be studied with detail and using simulations to understand their limitations before they are used in statistical analysis. The error estimation problem does not affect the One--Way ANOVA statistics, that reaches the highest performance to detect variations in the simulated differential lightcurves of quasars presented in this paper. In fact, the results present in this paper for the One--Way ANOVA test can be further improved by trialling different group sizes (\see \S\ref{sec:improving}). On the other hand, a discontinuous sampling based in the One--Way ANOVA methodology proves to be a powerful exploratory technique to detect microvariations. The relative loss of power of the discontinuous monitoring depends on the sampling frequency, but it is possible to tradeoff between this sampling frequency and an increase of the number of monitored objects switching target fields during the gaps between the groups of observations, resulting in a larger total number of detections of variability.

    The $F$ test for variances has less power than One--Way ANOVA and \chisq tests, but it is still a valid option to detect flux variations in AGNs. However, the \Cstat\ contains several misconceptions and cannot be considered a true statistical test.

\acknowledgments

    The author thanks Dr. Ignacio M\'{e}ndez and the anonymous referee for helpful suggestions.
    This work has been supported by CONACyT grant 50296 and has been
    partially funded by the Spanish MICINN under the Consolider-Ingenio 2010 Program grant CSD2006-00070: First Science with the GTC (\url{http://www.iac.es/consolider-ingenio-gtc}).

\appendix

\section{Tests description} \label{sec:test-desc}

    \subsection{$F$ test for sampled variances} \label{sec:F-test}

    Given two sample variances such as $s_Q^2$ for quasar differential lightcurve measurements and $s_*^2$ for the comparison star, the number of degrees of freedom for each sample, $\nu_Q$ and $\nu_*$ respectively, will be usually the same and equal to the number of measurements $N$ less one ($\nu = N - 1$). Then, if there are no differences in population variances, the sample variance ratio is distributed as the $F$ distribution with $\nu_Q$ and $\nu_*$ degrees of freedom:

    \begin{equation}\label{eq:ftest}
        F = \frac{s_Q^2}{s_*^2}
    \end{equation}

    This $F$ statistics is compared with the $F^{(\alpha)}_{\nu_Q,\nu_*}$ critical value, where $\alpha$ is the significance level set for the test, and $\nu_Q$ and $\nu_*$ the degrees of freedom of the quasar and the star of comparison samples. The smaller $\alpha$ value, the most improbable that the result is produced by chance. Thus a value $\alpha = 0.1\%)$ or 1\%, as assumed in this paper, roughly corresponds to a $3\sigma$ or a $2.6\sigma$ detection, respectively. If $F$ is larger than the critical value, the null hypothesis (no variability) is discarded. In this paper, the $F$ test has been performed at two significance levels (0.1\% and 1\%) to allow comparison with other statistical procedures. The respective critical values are  $F^{(0.001)}_{149,149} = 1.6651$ and $F^{(0.01)}_{149,149} = 1.4666$.

    A well known problem that may affect the outcomes of the $F$ test for variances is that it is non-robust to non-normality \citep[\S5.4]{Leh86}.

    \subsection{\chisq test for variance in a normal population} \label{sec:chi2-test}

    Given a number $N$ of observations of a source over a given period of time, these observations are supposed to be taken from a {\it population of possible observations} having a normal distribution. For this sample, the mean magnitude is $\overline{V}$, the $i$th observation yields a magnitude $V_i$ and the corresponding standard error $\sigma_i$. Then, the \chisq statistics is expressed by:

    \begin{equation}\label{eq:chisq}
    \chi^2 = \sum_{i=1}^N \frac{(V_i - \overline{V})^2}{\sigma_i^2}
    \end{equation}

    This statistics is compared against a critical value $\chi_{\alpha,\nu}^2$ obtained from the \chisq probability function, where $\alpha$ is the significance level and $\nu = N -1$ are the degrees of freedom. If $\chi^2 > \chi_{\alpha,\nu}^2$ the test indicates a larger than expected scattering of the data points (\ie evidence of variability). Each simulation comprises $N = 150$ observations, and thus $\nu = 149$. Therefore, the critical value adopted for this test is $\chi_{0.001,149}^2 = 208$.  

    \citet{Pic83}, \citet{Hei96}, \citet{Gop00}, \citet{And05} (\see references therein) and others follow the \chisq test proposed by \citet{Pen70} and \citet{Kes76}. This procedure uses a `weighted' average defined by:

    \begin{equation}
    \overline{V} = \frac{\sum_{i=1}^N V_i/\sigma_i^2}{\sum_{i=1}^N 1/\sigma_i^2}
    \end{equation}

    \noindent
    Note that $\sigma_i$ is the \emph{expected} error, \ie the error from considering photon noise from the source and sky, the CCD read-out and all possible non-systematic sources of error, some of them probably unknown in practice. As these individual errors are unknown, different estimates $s_i$ are used instead of $\sigma_i$. Thus, errors are often calculated from the usually underestimated value yielded by the IRAF reduction package, multiplied by a correction factor. This error rescalation is necessary because \chisq and other tests (but not ANOVA) assume that the distribution of the real errors is known.

    For example, \citet{Bac05} use a factor of 1.3; \citet{Sta04, Sta04a, Sta05} \citet{Gop03} and \citet[in the near infrared]{Gup08} use 1.5; \citet{Gar99} 1.73; and \citet{Gop95} 1.75. In fact, standard IRAF photometric packages do not take into account appropriate propagation of errors during the image processing. Particularly, \citet{Die98} have argued that these IRAF packages do not consider the possible spurious enhancement of the S/N ratio after the flat-field correction, due to changes in the sensibility across the detector (for example, sensibility near the borders may be lower than at the center), and that the error estimated directly from the corrected signal may be different from the original error. Note that this affects only the error estimation and not the measured level of the flux. Up to now, there is no astronomical reduction software that adequately deals with this problem, and implementing a solution may be not a trivial issue. ANOVA (\see \S\ref{sec:anova-test}) overcomes this problem by empirically measuring the dispersion within each group of observations (directly associated to the true S/N ratio), while data means for each group are conserved by flat-fielding the images.

    Usually, images of the object and the comparison stars (one or more) are recorded in the same CCD frames. However, combining several comparison stars to estimate the individual errors $\sigma_i$ in equation~\ref{eq:chisq} does not solve the problem. Here it is worth noting that $\chi_{\nu}^2$ with $\nu$ degrees of freedom distributes as $F_{\nu,\infty}$. Thus, even if $\sigma_i$ is obtained using several comparison stars to get a better estimate of the errors, the precision is still limited by the number of these stars, which probably are less than a dozen. In contrast, the \chisq test requires that each $\sigma_i$ would be calculated from all the possible measurements of an infinite population of stars. This issue is exemplified in \S\ref{sec:comparison}.



    Some authors substitute the individual errors by a common error $\sigma$ estimated from the dispersion in the comparison star data. This is the standard procedure used when performing the \Cstat\ (\see \S \ref{sec:C-test}). However, in differential photometry the number of images of the object and of the comparison stars is the same, and therefore the estimates for the standard deviation of their lightcurves have the same number of degrees of freedom. Then, the \chisq test is biased because it takes only into account the degrees of freedom of the estimation for the quasar. Therefore, the correct procedure is to consider the $F$ test for two sample variances as proposed by \citet{How88}. In conclusion, any procedure that cannot rely on  theoretically known (not estimated) error values, compromise the reliability of the \chisq test. Whenever the true errors are unknown,
    other statistical methodologies should be used.

    In the case of the simulations presented in this paper, all the parameters are controlled and the true errors can be effectively computed. Equation (\ref{eq:chisq}) has been used to compute the \chisq statistics. Note that data quality is ensured by the simulation design, thus there is no need to use weighted averages.

    \subsection{One-Way ANOVA test} \label{sec:anova-test}

    ANOVA tests are used to compare the means of a number of samples.
    Due to the {\it Central Limit Theorem}, no matter what the shape of the original distribution is, the sampling distribution of the mean approaches a normal distribution. Therefore, tests to compare the means ($t$-test and ANOVA) are robuster than their counterparts to compare variances (\chisq and $F$ tests). This offsets the problem of the non-robustness of the $F$ test.

    One--Way ANOVA has been applied by \citet{Die98} to investigate the variability in the lightcurves of quasars. The methodology consisted in measuring $k$ groups of $n_j = 5$, one after the other, short (1\,min) observations. The $k$ groups are ideally separated by $20-30$\,min. Larger time lags are common, affecting the time resolution but not the statistical significance of the test.

    For the mathematical description of the One--Way ANOVA test, if $y_{ij}$ represents the $i$th observation (with $i=1,\,2,\,...n_j$) on the $j$th group (with $j=1,\,2,\,...k$), the linear model describing every observation is:

    \begin{equation}\label{eq:anova_linear_model}
    y_{ij} = \overline{y} + g_j + \varepsilon_{ij},
    \end{equation}

    \noindent
    where $\overline{y}$ represents the mean of the whole data set, $g_j = \overline{y}_j - \overline{y}$ the between--groups deviation, and  $\varepsilon_{ij} = y_{ij} - y_i$ the within--groups deviation, also called residual or measurement error. The size of the data set will be $N = \sum_{i=1}^k n_j$. If the number of observation in the groups $n_j$ is constant, $N = k \times n_j$.

    As commented previously, ANOVA tests whether the means of the groups are equal. The condition tested or \emph{null hypothesis} is that the means of the different groups are equal. If the test yields a probability smaller than the adopted significance level $\alpha$, the null hypothesis will be rejected and the alternate hypothesis (at least one group mean is different from the others), will be accepted. The alternate hypothesis in this case implies detection of variability in the quasar lightcurve.

    From equation~(\ref{eq:anova_linear_model}), the total sample variation can be separated into variations \emph{between} and \emph{within} groups

    \begin{equation}\label{eq:anova_terms}
        \sum_{j=1}^k \sum_{i=1}^{n_j} (y_{ij} - \overline{y})^2 = \sum_{j=1}^k (y_{j} - \overline{y})^2 + \sum_{j=1}^k \sum_{i=1}^{n_j} (y_{ij} - \overline{y}_j)^2,
    \end{equation}

    \noindent
    where the term in the left side describes the total deviations of the data with respect to the mean. The firs term in the right side of the equation represents the total variation between groups, and the last term the total errors. Equation~(\ref{eq:anova_terms}) can be shortened to:

    \begin{equation}
    SS_T = SS_G + SS_R
    \end{equation}

    \noindent
    where $SS_T$ stands for the total \emph{sum of squares}. Similarly $SS_G$ and $SS_R$ stand for group sum of squares and residual sum of squares.

    Whenever the null hypothesis is true, the $k$ groups of sampled data will be normally and independently distributed, with mean $\mu$ and variance $\sigma^2$. Then, the statistics:

    \begin{equation}\label{eq:fstat}
         F = \frac{SS_G / (k-1)}{SS_R / (N-k)} = \frac{MS_G}{MS_R},
    \end{equation}

    \noindent
    corresponds to the $F$ distribution with $\nu_1=k-1$ and $\nu_2=N-k$ degrees of freedom. The \emph{pseudo} variances $MS_G$ and $MS_R$ are mean estimates for the variations between groups and residuals, respectively. For a certain significance level $\alpha$, if $F$ exceeds the critical value $F^{(\alpha)}_{\nu_1,\nu_2}$ the null hypothesis will be rejected.

    The $F$ critical values employed in this paper for the ANOVA tests are $F^{(0.001)}_{29,120} = 2.2819$ and $F^{(0.001)}_{9,40} = 4.0243$ for the full and the 30\,min lags sample simulations, respectively.

    \subsection{\Cstat} \label{sec:C-test}

    \Cstat\ was first employed by \citet{Jan97} and generalized by \citet{Rom99}. The statistical parameter used is

    \begin{equation}\label{eq:cstat}
        C = \frac{\sigma_T}{\sigma}
    \end{equation}

    \noindent
    where $\sigma_T$ and $\sigma$ are the standard deviation of the quasar and the comparison star differential lightcurves. The adopted variability criterion requires that $C \geq 2.576$ which corresponds to a 99\% confidence level, or 1\% significance level following the notation in this paper.

    There are two pitfalls with this criterion. First, the critical value 2.576 corresponds to the 1\% significance level of the normal distribution for a \emph{two-sided} test, rather than a \emph{one-sided} comparison. The two-sided test would be relevant to test that the dispersion in the quasar lightcurve may be both, larger or smaller, than the dispersion of the comparison star lightcurve. Note also that \Cstat\ would always be positive, and that its expected value when $\sigma_T = \sigma$ is centered around 1, rather than 0 as would be expected in the case of a fair normal distribution.

    Another pitfall, probably the most important, is that you cannot compare two standard deviations using the normal distribution. In fact, it is unfeasible to use standard deviations for most calculations because they are not lineal statistical operators (for example, given two independent random variables $A$ and $B$ with standard deviations $\sigma_A$ and $\sigma_B$, respectively, the standard deviation of the sum $A+B$ is not $\sigma_A + \sigma_B$). That is the reason because you have to use variances instead; variances are the second moments of the statistical distributions and therefore lineal operators (in the previous example, the variance of the sum $A+B$ is $\sigma_A^2 + \sigma_B^2$).

    If we draw all possible samples of a given size $N$ from a normally distributed population and compute the variances of all those samples, we will obtain a distribution of sample variances that starts with $s^2 = 0$ and have a mean of $\sigma^2$.
    Thus, even if the distribution of all possible sample means drawn from a normally distributed population will be approximately symmetrical (as a consequence of the Central Limit Theorem), an equivalent distribution of sample variances will not approximate symmetry, but will be distributed as \chisq with $N-1$ degrees of freedom (this result is a consequence of the Cochran's Theorem for the sum of squares of linear combinations of a set of independent standard normally distributed random variables):

    \begin{equation} \label{eq:var_dist}
    \chi^2 = (N-1) \frac{s^2}{\sigma^2}
    \end{equation}

    One important outcome of this equation is that the the shape of the \chisq distribution will be different for different sample sizes. Thus, the dispersion of the distribution of sample variances will depend on how many degrees of freedom has our estimate. Note that this dependence of the variance dispersion on the number of degrees of freedom transmits to the standard deviation, although the \chisq statistics does not apply in this case.


    Using the $F$ statistics it is easy to calculate the `real' significance of the \Cstat. If the critical value from equation~(\ref{eq:cstat}) is 2.576, the critical $F$ value is $2.576^2 = 6.636$. For $F$ with (20,20) and (50,50) degrees of freedom, the significance level of the test will be $4.4 \times 10^{-5}$ and $2.1 \times 10^{-10}$, respectively; \ie much less than the $10^{-2}$ value considered by \citet{Jan97} and \citet{Rom99}. Note that \citeauthor{Rom99} report that variations in radio loud quasars occur more often than in radio quiet quasars, but most sources in their radio loud sample are strongly variable BL\,Lac objects that can easily reach high significance levels of microvariability detection.

    To summarize this discussion, the critical values for the \Cstat\ criterion are wrongly established, and equation~(\ref{eq:cstat}) does not describe a normal distributed variable because it is neither properly centered such as the mean expected value is zero, nor does the independent sample size parameter $\sigma$ represent an unbiased measurement of the dispersion of the sample standard deviation. The square value of the \Cstat\ can be used to perform an unbiased $F$ test if the degrees of freedom used for the $\sigma_T$ and $\sigma$ estimates are provided.

\section{Improving the ANOVA power} \label{sec:improving}

    The power of any statistical test to study variability during a given time interval depends on the number of observations recorded to improve the time resolution, and the measurement precision to improve the amplitude resolution. In practise, a compromise is attained to hold reasonable resolutions for both factors \emph{before} actually observing the target. Would not be useful to have the possibility of swapping between the time and amplitude resolutions \emph{after} the observations? Binning data might help, but it is shown in \S\ref{sec:discussion} that it does not work for a \chisq methodology because the loss of degrees of freedom.

    In the ANOVA methodology, if the target has been monitored discontinuously, as in the case of ANOVA with 30\,min lags, the binning is fixed and consequently the time and amplitude resolutions cannot be exchanged either. But if the monitoring was continuous, it is possible to try out different bin sizes to improve the amplitude resolution at the expense of the time resolution. In this paper, a binning of 5 observations was used because it was \emph{a priori} reasonable value for time resolution and allowed direct comparison with the ANOVA with 30\,min lags strategy. Now, different bins of $n$ observations will be considered to achieve the maximum power of the ANOVA test for the lightcurve simulations considered in this paper. The bin size $n$ will always be a divisor of the total number $N=150$ of observations of a lightcurve simulation.

    Table~\ref{tab:binnings} shows the ANOVA results for the Gaussian peak variations considering different bin sizes strategies. Column (1) indicates the bin size $n$; column (2) shows the number of degrees of freedom $\nu_1$ for the $k$ groups ($\nu_1 = k - 1$); column (3) displays the number of degrees of freedom $\nu_2$ for the residuals ($\nu_2 = N - k$); column (4) shows the F critical value for a significance level $\alpha = 0.1\%$, and $\nu_1$ and $\nu_2$ degrees of freedom ($F^{(0.001)}_{\nu_1,\nu_2}$); column (5) indicates the number of Type\,I errors ; column (6) shows the number of detections of variability; and column (7) displays the relative power for the ANOVA tests normalized to the results for $n=15$.

\begin{table}[t]
 \center{
 \begin{minipage}{0.55\textwidth}
  \caption{ANOVA performance for different bin sizes.} \label{tab:binnings}
  \begin{tabular}{rrrr@{.}lrrr}
  \tableline\tableline
         &         &         &  \multicolumn{2}{c}{} & \multicolumn{2}{c}{Gaussian peak variation} & Relative \\
         \cline{6-7}
     $n$ & $\nu_1$ & $\nu_2$ & \multicolumn{2}{c}{$F^{(0.001)}_{\nu_1,\nu_2}$} & Type\,I   & Detections  &  Power \phantom{i} \\
 \tableline
      2 & 74 &  75 &  2      & 0657    & 2         & 1178       & 0.48 \\
      3 & 49 & 100 &  2      & 0835    & 5         & 1713       & 0.69 \\
      5 & 29 & 120 &  2      & 2819    & 3         & 2187       & 0.88 \\
      6 & 24 & 125 &  2      & 3907    & 6         & 2268       & 0.92 \\
     10 & 14 & 135 &  2      & 8199    & 4         & 2444       & 0.99 \\
     15 &  9 & 140 &  3      & 3374    & 1      & \textbf{2478} & 1.00 \\
     25 &  5 & 144 &  4      & 3617    & 3         & 2285       & 0.92 \\
     30 &  4 & 145 &  4      & 8882    & 7         & 2143       & 0.86 \\
     50 &  2 & 147 &  7      & 2428    & 5         & 1996       & 0.81 \\
     75 &  1 & 148 & 11      & 2727    & 4         &  430       & 0.17 \\
 \tableline
\end{tabular}
\end{minipage} }
\end{table}

    A careful inspection of Table~\ref{tab:binnings} shows that, for the simulations presented in this paper, the test power to detect variations increases with the bin size until reaching a maximum of 2478 detections for $n=15$ (noted in boldface in Table~\ref{tab:binnings}), and then decreases for larger bin sizes. The smooth overall detection tendency shows that the differences in power for these tests with different bin sizes are not an artifact. For $n=5$ the relative power is almost 0.9, and all the bins between $n=5$ and 50 have relative power larger than 0.8. These results show that the ANOVA bin size is not a fine tuning parameter, and that binning the data in groups of 5 observations is, also from the statistical point of view, a reasonable choice.

    The smallest bin sizes improve the time resolutions, but the detections are biased towards large amplitude variations. On the contrary, the largest bin sizes can detect small variations, but the time resolution is degraded and fast variations are not detected. Thus, the tests for $n=5$ detect 65 variations with time FWHM up to 13\,min, while the tests for $n=15$ detect only 37. But the percentages of detections for an amplitude of variation of approximately 0.01\,mag are around 10\% and 20\% for the $n=5$ and $n=15$ tests, respectively.

    In the case of linear variations, or more generally monotonic variation, the results are simpler. As the duration of the variability event is not an issue in this case, the only factor that matters is the amplitude resolution. Therefore, the tests that have more power are those with only two groups, one for each half of the lightcurve, that corresponds to $n=75$, $\nu_1=1$ and $\nu_2=148$. The actual number of detections for this test is 4080, instead of 3043 detections that were obtained for $n=5$.

    These results are easily translated to the case of analyzing a single quasar lightcurve. The researcher does not know in advance the specific characteristics of a possible microvariability event. But planning the observations to be analyzed using an ANOVA experimental design, the astronomer can enhance the temporal resolution of the observations to detect very fast large amplitude variations, and still conserve the amplitude resolution to detect longer timescale low amplitude variations. Of course, this has also a cost. The final significance level $\alpha_f$ when performing a set of $n_t$ statistical tests is different from the significance level $\alpha_t$ for the individual tests. There are a number of adjustments applied in the statistical literature for multiple tests, the most common is the Bonferroni correction:
    \[
    \alpha_t = \alpha_f / n_t
    \]
    If the set of tests performed are the same as in the case presented above for the simulations, $n_t = 10$ and $\alpha_t = 0.1\%$, the final significance level will be $\alpha_f = 1\%$. Similarly, the significance level of the individual tests could be set to $\alpha_t=0.01\%$ to reach a final value of $\alpha_f=0.1\%)$. However, the results shown in Table~\ref{tab:binnings} suggest that performing test for bins of size 5, 15 and 30 may be enough to ensure detection on a wide range of timescales and amplitudes. If only three tests are performed, the significance levels are $\alpha_f = 0.3\%$ (if $\alpha_t$ is set at 0.1\%) or $\alpha_t = 0.03\%$ (if $\alpha_f$ is maintained at 0.1\%).



\end{document}